# Coordinated *PV re-phasing*: a novel method to maximize renewable energy integration in LV networks by mitigating network unbalances

W.G. Chaminda Bandara[a], G.M.R.I. Godaliyadda[a], M.P.B. Ekanayake[a], J.B. Ekanayake*[,ab]

[a]*Department of Electrical and Electronic Engineering, University of Peradeniya, Peradeniya (20400), Sri Lanka.*
[b]*School of Engineering, Cardiff University, The Parade, Cardiff CF24 3AA, United Kingdom.*

## Highlights

- Maximize renewable energy penetration through coordinated re-phasing of solar PV.
- An automatic PV re-phasing switch that can connect to single-phase PV inverters.
- A modified DBFOA to determine the optimum phase combination of PV systems.
- The key mechanisms of DBFOA are modified to cater to PV re-phasing problem.
- A contextually optimized initializer to improve the convergence speed of DBFOA.

## Abstract

As combating climate change has become a top priority and as many countries are taking steps to make their power generation sustainable, there is a marked increase in the use of renewable energy sources (RESs) for electricity generation. Among these RESs, solar photovoltaics (PV) is one of the most popular sources of energy connected to LV distribution networks. With the greater integration of solar PV into LV distribution networks, utility providers impose caps to solar penetration in order to operate their network safely and within acceptable norms. One parameter that restricts solar PV penetration is unbalances created by loads and single-phase rooftop schemes connected to LV distribution grids. In this paper, a novel method is proposed to mitigate voltage unbalance in LV distribution grids by optimally re-phasing grid-connected rooftop PV systems. A modified version of the discrete bacterial foraging optimization algorithm (DBFOA) is introduced as the optimization technique to minimize the overall voltage unbalance of the network as the objective function, subjected to various network and operating parameters. The impact of utilizing the proposed PV re-phasing technique as opposed to a fixed phase configuration are compared based on overall voltage unbalance, which was observed hourly throughout the day. The case studies show that the proposed approach can significantly mitigate the overall voltage unbalance during the daytime and can facilitate to increase the usable PV capacity of the considered network by 77%.

***Keywords:*** Renewable energy integration, Rooftop solar PV, PV re-phasing, network unbalance, LV distribution networks, Bacterial foraging optimization.

* Corresponding author.
*Email addresses:* chaminda.bandara@eng.pdn.ac.lk (W.G. Chaminda Bandara), roshangodd@ee.pdn.ac.lk (G.M.R.I. Godaliyadda), mpb.ekanayake@ee.pdn.ac.lk (M.P.B. Ekanayake),
jbe@ee.pdn.ac.lk, ekanayakej@cardiff.ac.uk (J.B. Ekanayake*)

# 1. Introduction

With the de-carbonization agenda of many countries, the integration of renewable energy sources (RES) such as solar and wind energy into the power system has increased considerably. Among these RES, photovoltaic solar energy (PV) is one of the fastest-growing renewable energy sources with an annual growth rate of 35 to 40% [1]. At the end of 2019, the global cumulative solar energy capacity stood at 580 GW. These installations come in three forms: ground-mounted PV (GPV) [2, 3], floating PV (FPV) [4, 5], and rooftop PV [6, 7]. As GPV and FPV use dedicated feeders for connection to the network, there is no locals impact on the network, whereas the random installation of small-size rooftop PV systems in LV networks causes a voltage unbalance. The unbalance in phase voltages and the resulting flow of large neutral currents can increase the distribution and transformer losses due to overheating and can overload the neutral conductor [8, 9]. Due to these consequences of voltage unbalance, utility providers limit the usable PV capacity that they can accommodate for LV networks [10–12]. Therefore, an effective solution is needed to improve the future integration of solar PV sources into LV networks. How to effectively reduce the unbalance is a long-standing question, and much effort has been devoted to answering this question in the past decade.

In the literature, many techniques have been proposed to minimize the unbalance. These techniques can be divided mainly into two categories [13]. The first category mitigates the network unbalance by using neutral current compensation devices such as passive harmonic filters/ specially designed transformers (e.g. synchronous machines as filters [14], T-connected transformer [15], star-hexagon transformer [16], zigzag transformer with single-phase series/shunt active power filter (APF) [17], and star-delta transformer with single-phase half-bridge PWM [18]), and specially designed active power filters (e.g. H-bridge shunt APF [19], three-phase four-wire capacitor midpoint APF [20] and three-phase four-wire four-leg APF topology [21]). The second category is based on distribution network reconfiguration techniques and can be divided into two: distribution feeder reconfiguration (DFR) and phase balancing. It is important to note that both DFR and phase balancing techniques use non-linear, non-differentiable, highly combinatorial, and constrained optimization algorithms to find the optimal solution [22].

The Distribution Feeder Reconfiguration (DFR) technique optimizes the open or closed state of sectionalizing switches and tie switches to transfer the loads from overloaded feeders to the lightly loaded feeders to minimize desired objective functions (e.g., voltage unbalance, load unbalance, power loss, etc.) while preserving the radial configuration of the LV distribution system [23]. Further, many researchers have used different optimization techniques to mitigate voltage unbalance and power loss using DFR techniques. These include heuristic search [24], ant colony optimization [25, 26], genetic algorithm [24, 27–29], incremental algorithm [30], fuzzy approach [31, 32], colored Petri net algorithm [33, 34], second-order cone programming [35], mixed-integer linear programming [36, 37], and hybrid bacterial foraging - spiral dynamic [38] algorithms. The time-varying nature of loads and distributed generators, and their uneven distribution in the network cause LV feeders to often become unbalanced. Although the DFR techniques can only mitigate unbalance at the system level it cannot mitigate phase unbalance at the feeder level [13]. Hence, phase balancing techniques have been proposed to mitigate feeder level unbalance.

The phase balancing technique can be implemented in two ways: (1) load re-sequencing and (2) load re-phasing. In the load re-sequencing technique, the phase sequence at each busbar is re-sequenced to their optimal combination. To avoid reverse operation of inductive loads, the positive and negative phase sequences are only taken into account [39]. In the load re-phasing technique, the loads from the overloaded phases are transferred to the lightly loaded phases by analyzing the current or power difference between the phases. To identify the optimum phase sequence for the three-phase loads and optimum phase combination for the single-phase loads, different optimization techniques have been proposed. Examples of these include, heuristic search [40], mixed integer programming [41], fuzzy logic and combinatorial optimization [42], particle swarm optimization [43], bacterial foraging – particle swarm optimization [43, 44], simulated annealing [45] and genetic algorithm [46]. However, these techniques have been tested on small LV networks with a few loads and have not been implemented in large LV networks due to the high computational time they use to identify the optimum solution. Also, the aforementioned unbalance mitigation techniques raise many concerns when they implemented in practical networks such as:

- *high initial cost* due to the fact that load switches need to be installed between the phases, at each end of the client [13],
- *several other indirect costs* such as the cost of customer interruption, the cost of customer reliability, etc. [47], and
- *possible harm or damage* to the customer equipment at the time of re-phasing.

Considering the above limitations of load re-phasing, in this paper, a PV re-phasing technique is proposed that only relies on the rooftop solar systems to minimize the network unbalance. Therefore, the re-phasing switches need to be installed only at the connection point of each rooftop solar system and require much fewer re-phasing switches per network as compared to load/feeder re-phasing techniques. Due to this low capital requirement for the implementation of the proposed PV re-phasing technique, it is ideal for the large-scale deployment in LV networks to minimize the network unbalance. Also, the proposed PV re-phasing technique has no impact on supply reliability and requires minimum or no customer interruption. Therefore, the proposed PV re-phasing technique is a more economical and effective way to minimize the network unbalance, and thereby facilitate to improve the integration of clean and renewable solar energy into the LV networks. The main contributions of this paper are listed as follows:

- ***A novel strategy to minimize unbalance in LV networks based on automatic re-phasing of grid-connected rooftop PV systems.*** The grid-connected rooftop solar systems are periodically re-phased to their optimal phase combination at pre-selected time intervals to minimize the system unbalance. The proposed PV re-phasing technique can maintain the voltage unbalance well below the 1% threshold line while simultaneously maintaining the phase voltages within their acceptable limits. This will help utility providers to allow more rooftop solar systems into the network without bothering about network unbalance. The case studies demonstrate that the proposed PV re-phasing strategy can improve the usable PV capacity of the considered network by 77%.
- ***A PV rephasing switch is proposed to perform automatic rephasing of grid-connected single-phase PV systems.*** Since the re-phasing switch only re-phases rooftop PV systems, not the loads or the feeders, it will not have any impact on supply reliability. Additionally, no customer interruption is required at the time of PV re-phasing.

- ***A discrete bacterial foraging optimization algorithm (DBFOA) was introduced to determine the optimal phase combination of grid-connected single-phase PV systems.*** The proposed DBFOA improves upon the classical BFOA by modifying the principal mechanisms of the classical method to specifically catered to the PV re-phasing problem, thereby increasing both convergence speed and accuracy. The proposed bacterial foraging optimization algorithm determines the optimal phase combination for rooftop PV systems such that it will minimize the violations of voltage unbalance and phase voltage magnitudes.

The rest of the paper is organized as follows. Section 2 describes the structure of the proposed re-phasing switch and the overall operating mechanism of the PV re-phasing technique. Section 3 formulates the PV re-phasing problem. Section 4 describes the principal mechanisms of the proposed bacterial foraging optimization algorithm and the implementation details (i.e. the flow chart and the pseudo-code), as well as the simulation results, are given in section 5. Finally, section 6 presents our conclusions and future work.

# 2. PV re-phasing arrangement

In this section, the structure of the re-phasing switch and the operating mechanism of automatic PV re-phasing in LV distribution grids are explained.

## 2.1. Structure of the PV re-phasing switch

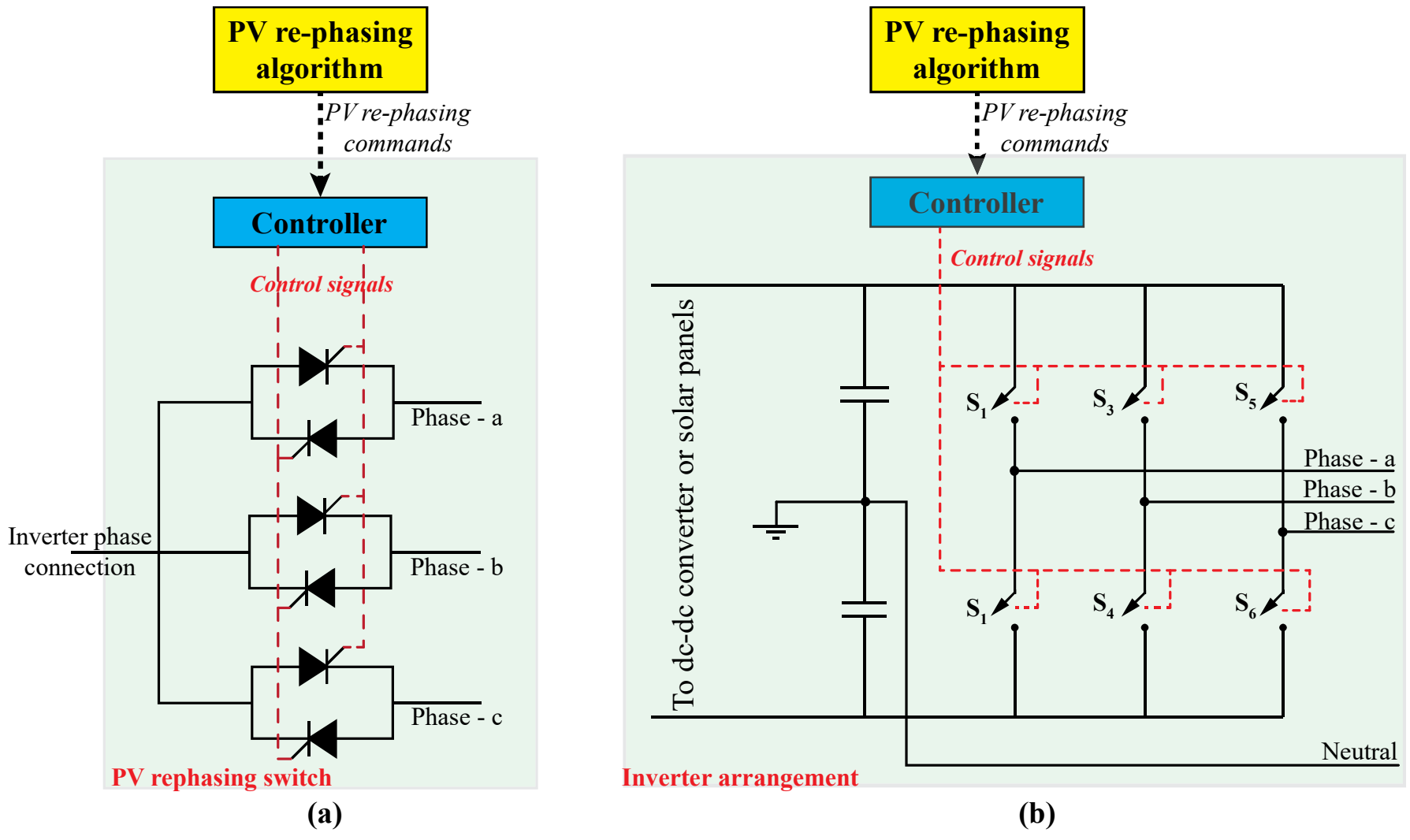

Figure 1: (a)-Thyristor switch, (b)-Inverter arrangement

Figure 1-(a) shows a schematic of a PV rephasing switch that can be connected to the output of the single-phase inverter. As can be seen, switching between phases can be achieved by blocking the already conducting pair of thyristors (or trial) and turning the pair of thyristors that are connected to the phase to which output should be connected. In order to avoid any circulating

current between phases, a dead band should be introduced between the blocking signal and the turning on signal. However, this will eventually course the PV inverter to take the start-up mode, thus, introducing an interruption of a few minutes (less than 3 min). The inverter shown in Figure 1-(b) can be used to prevent such a transient. In this arrangement, a half-bridge inverter is used to convert dc into ac. If the output needs to be connected to Phase - a then switches $S_1$ and $S_2$ are operated in a complementary manner using a PWM switching pattern and all the other switches are blocked. If the output needs to be connected to Phase - b then $S_1$ and $S_2$ will be blocked and $S_3$ and $S_4$ will be turned on. As these switching transients take nanosecond level times, it can be used as a rapid rephasing arrangement but with extra cost for the two switching arms. The same arrangement can be introduced for full-bridge single-phase inverters.

## 2.2. The architecture of the automatic PV re-phasing arrangement

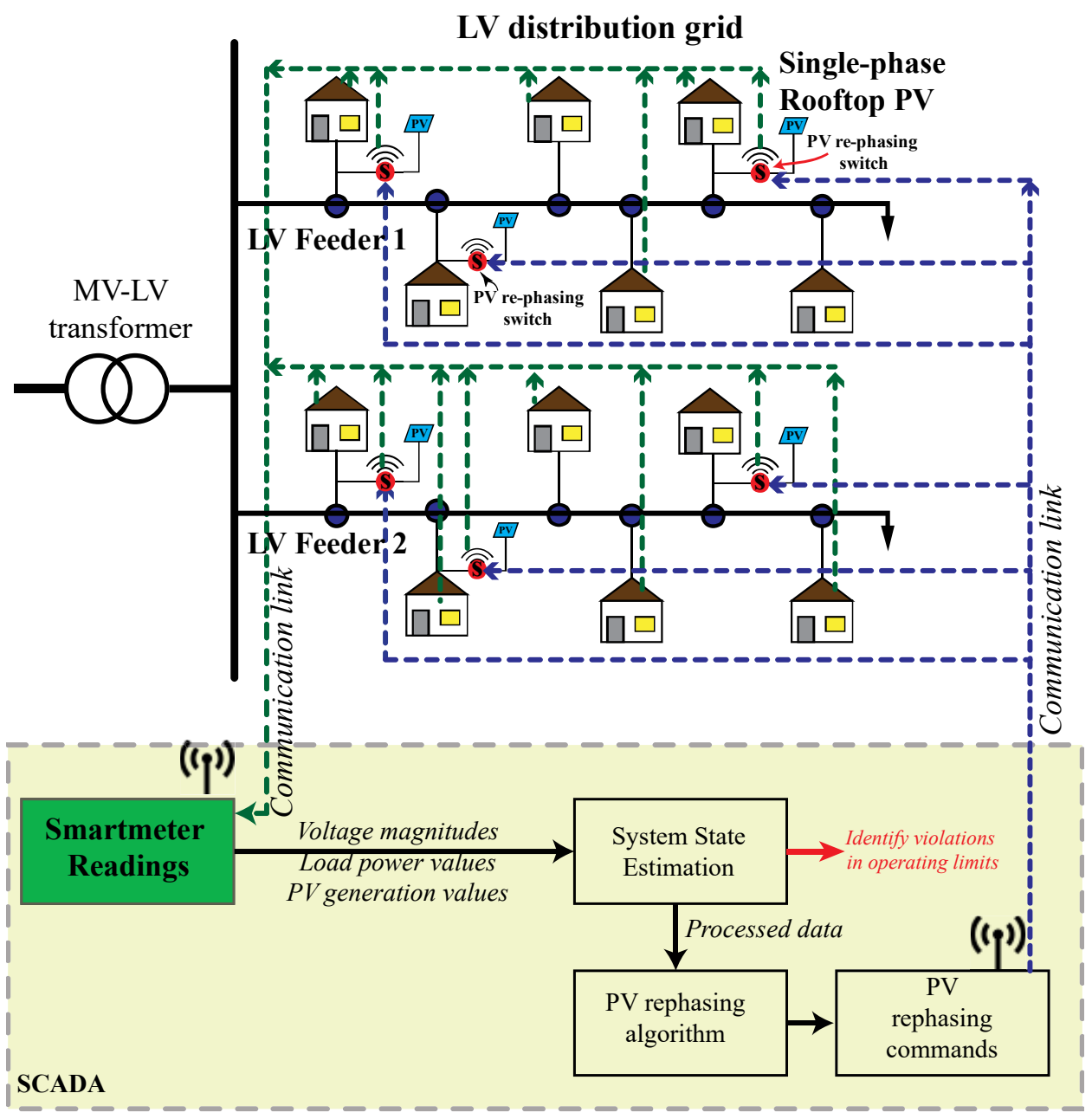

Figure 2: Schematic of the automatic PV re-phasing arrangement

Figure 2 shows the operating mechanism of the proposed PV re-phasing strategy. The necessary data such as PV generations and load demands are measured from smart meters and transmit to the supervisory control and data acquisition unit (SCADA). Typically, these smart meter measurements are subjected to different types of systematic, random, and communication errors [48, 49]. Therefore, in the next step, state estimation is performed to detect the presence of bad data. These preprocessed smart meter data are sent to the PV re-phasing algorithm to determine the optimal phase combination of grid-connected PV systems such that the overall voltage unbalance of the network is minimized. Finally, the re-phasing program transmits the required phase changes of PV systems to the SCADA system. Then, SCADA sends required re-phasing commands to the installed PV re-phasing switches and re-phasing operation is carried out.

## 3. Problem formulation

The aim of this work is to develop a strategy to minimize the overall voltage unbalance of the network such that the power quality and the reliability of the distribution system can be improved. The objective of this optimization problem can be expressed as the minimization of the mean voltage unbalance factor ($\overline{VUF}$) of the network as in,

$$\overline{VUF} = \frac{1}{N}\sum_{n=1}^{N} VUF_n \quad (\mathbf{1})$$

where, $VUF_n$ is the voltage unbalance factor at $n$-th busbar and $N$ is the total number of busbars in the network.

***Subjected to the constraints:***

1. Voltage unbalance at each busbar ($VUF_n$) must be strictly below the specified maximum unbalance level ($VUF_{max}$):
$$VUF_n \le VUF_{max} \quad (\mathbf{2})$$
for $n = 1,2,3,\dots,N$.
2. Phase voltage magnitudes ($V_n^a, V_n^b,$ and $V_n^c$ ) must strictly between the upper ($V_{max}$) and lower ($V_{min}$) limits:
$$V_{min} \le V_n^a, V_n^b, V_n^c \le V_{max} \quad (\mathbf{3})$$
for $n = 1,2,3,\dots,N$, where, $V_n^a$, $V_n^b$ and $V_n^c$ are the voltage magnitudes of $a$, $b$, and $c$ phases at $n$-th busbar, respectively.

Equation **(1)** corresponds to the objective function to be minimized and represents the overall voltage unbalance ($\overline{VUF}$) of the distribution network. The inequality in **(2)** considers a constraint for voltage unbalance factor and ensures individual voltage unbalance factors ($VUF_n$ for $n = 1,2,..,N$) are below the specified maximum value, $VUF_{max}$. The inequality in **(3)** deals with the constraints for voltage magnitudes. It ensures the phase voltages ($V_n^a, V_n^b,$ and $V_n^c$ ) fall within the acceptable voltage limits (lower limit $V_{min}$ and upper limit $V_{max}$). In this study, $V_{min}$ was considered as 0.94 pu and $V_{max}$ was considered as 1.06 pu. In other words, **(2)** and **(3)** define the feasible regions for voltage unbalance ($VUF_n$) and phase voltage magnitudes ($V_n^a, V_n^b, V_n^c$), respectively.

In order to minimize **(1)** while simultaneously satisfying the constraints **(2)** and **(3)**, penalty functions were introduced. The main idea of these penalty functions is that an optimal PV configuration (i.e. the optimal solution) requires that constraints be active so that this optimal solution lies in the feasible regions for voltage unbalance and phase voltage magnitudes. To ensure this, a penalty is applied to possible solutions when constraints are not satisfied. Therefore, the aforementioned optimization problem was reformulated as the minimization of the penalized objective function, $J(\boldsymbol{x})$, given by,

$$J(\boldsymbol{x}) = \overline{VUF} + k_1 \sum_{n=1}^{n=N} \mu_{VUF_n} + k_2 \left( \sum_{n=1}^{n=N} \mu_{V_n^a} + \sum_{n=1}^{n=N} \mu_{V_n^b} + \sum_{n=1}^{n=N} \mu_{V_n^c} \right) \qquad \mathbf{(4)}$$

where,

the penalty function for voltage unbalance ($\mu_{VUF_n}$) is given by:

$$\mu_{VUF_n} = \begin{cases} VUF_n - VUF_{max} \; ; \text{ when } VUF_n > VUF_{max} \\ 0 \qquad\qquad\qquad\; ; \text{ when } VUF_n \leq VUF_{max} \end{cases} \qquad \text{for } n = 1, \dots, N,$$

the penalty function for voltage magnitudes of phase A ($\mu_{V_n^a}$) is given by:

$$\mu_{V_n^a} = \begin{cases} |V_n^a - V_{min}| \; ; \text{when } V_n^a < V_{min} \\ \qquad 0 \quad ; \text{when } V_{min} \leq V_n^a \leq V_{max} \\ V_n^a - V_{max} \quad ; \text{when } V_n^a > V_{max} \end{cases} \qquad \text{for } n = 1, \dots, N,$$

the penalty function for voltage magnitudes of phase B ($\mu_{V_n^b}$) is given by:

$$\mu_{V_n^b} = \begin{cases} |V_n^b - V_{min}| \; ; \text{when } V_n^b < V_{min} \\ \qquad 0 \quad ; \text{when } V_{min} \leq V_n^b \leq V_{max} \\ V_n^b - V_{max} \quad ; \text{when } V_n^b > V_{max} \end{cases} \qquad \text{for } n = 1, \dots, N,$$

the penalty function for voltage magnitudes of phase C ($\mu_{V_n^c}$) is given by:

$$\mu_{V_n^c} = \begin{cases} |V_n^c - V_{min}| \; ; \text{when } V_n^c < V_{min} \\ \qquad 0 \quad ; \text{when } V_{min} \leq V_n^c \leq V_{max} \\ V_n^c - V_{max} \quad ; \text{when } V_n^c > V_{max} \end{cases} \qquad \text{for } n = 1, \dots, N,$$

$\boldsymbol{x}$ is the PV configuration vector, and $k_1$ and $k_2$ are the constant imposed on the penalty functions. The graphical illustrations of these penalty functions are shown in Figure 3.

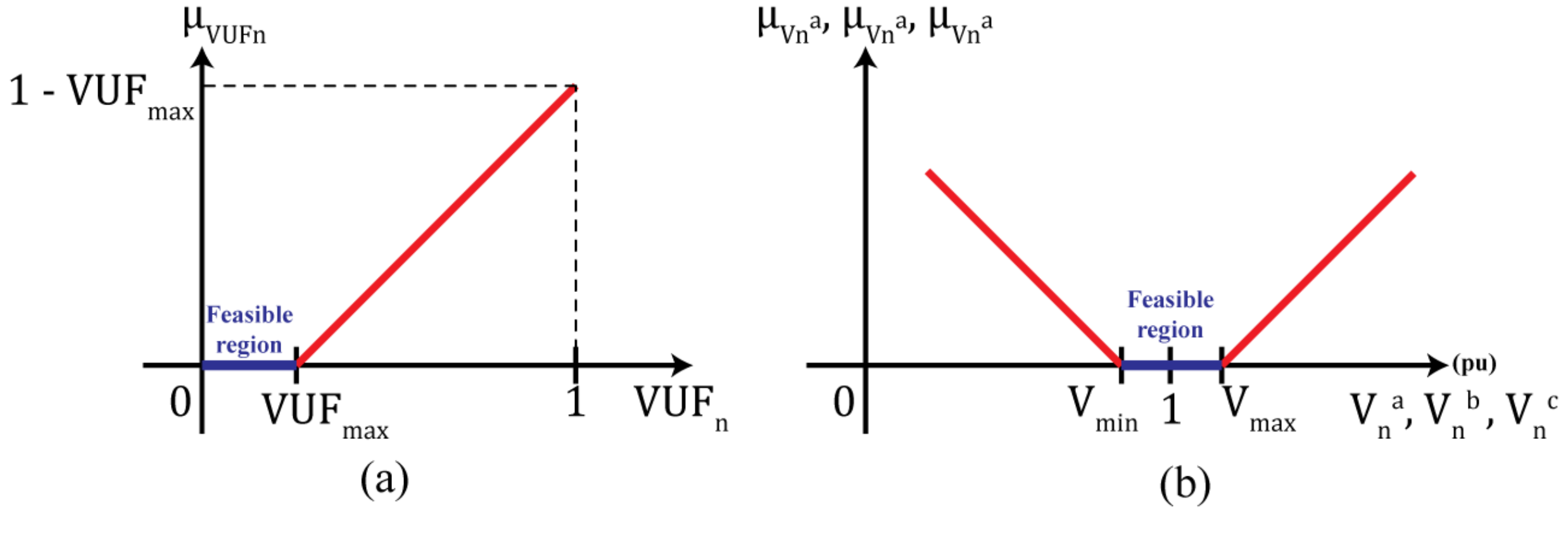

Figure 3: Penalty functions for (a) voltage unbalance and (b) phase voltage magnitudes.

The phase combination of grid-connected PV systems at a given time is represented by the *PV configuration vector*, $\boldsymbol{x}$. Therefore, for a network having $N_{pv}$ number of grid-connected PV systems, the PV configuration vector $\boldsymbol{x}$ consists of $N_{pv}$ number of phase entries where each phase entry corresponds to the phase of a grid-connected PV system in the network. Hence, each element ($x_m$) in the PV configuration vector-$\boldsymbol{x}$ can take one of the three phases ($i.e. \boldsymbol{x}_m \in \mathbb{P}, \mathbb{P} = \{a, b, c\}$). The format of the PV configuration vector is illustrated in Figure 4 with an example PV combination.

Format of the PV configuration vector $\boldsymbol{x} \in \mathbb{P}^{N_{pv}}$

| | | | | | | | | | |
|---|---|---|---|---|---|---|---|---|---|
| Phase of $PV_1$ | Phase of $PV_2$ | Phase of $PV_3$ | … | … | Phase of $PV_m$ | … | … | … | Phase of $PV_{N_{pv}}$ |

Example PV configuration vector $\boldsymbol{x}^i$

| | | | | | | | | | |
|---|---|---|---|---|---|---|---|---|---|
| Phase a | Phase b | Phase c | … | … | $x_m \in \mathbb{P}$ | … | … | … | Phase a |

Figure 4: The format of the PV configuration vector $\boldsymbol{x}$

It is important to note that there are different definitions available for the voltage unbalance factor; in this paper, the IEC definition [50] was used. In the IEC definition (IEC TR 61000-3-14:2011), the voltage unbalance factor is calculated as the ratio of negative sequence voltage component to the positive sequence voltage component and can be expressed as follows [50]:

$$VUF_n = \frac{V_n^-}{V_n^+} = \frac{\text{Negative sequence voltage component at } n^{\text{th}} \text{ busbar}}{\text{Positive sequence voltage component at } n^{\text{th}} \text{ busbar}} \times 100\% \qquad (\mathbf{5})$$

The three-phase sequence voltage components were obtained by the symmetrical transformation. The steps for the computation of three-phase sequence voltage components from three-phase voltages can be found in [39].

# 4. Bacterial Foraging Optimization

Bacterial foraging optimization algorithm (BFOA) is a smart optimization technique that has proven to be very effective in search domains having several dimensions. BFOA is inspired by the social foraging behavior of *E. coli* bacteria. The underlying biology behind the foraging strategy of *E. coli* is emulated and used as a simple optimization algorithm [51, 52]. In this paper, a discrete and adaptive version of BFOA is introduced to solve the PV re-phasing problem.

## 4.1. Concept of BFOA

During the foraging period, real bacteria achieve their motion with the help of their tensile flagella. Using these tensile flagella, bacteria can perform two basic motion types called tumble and swim. In the classical BFOA, the bacteria undergo *chemotaxis*, where they like to move towards nutrient gradient while avoiding the noxious environments. When they get enough food, they increase their length and under suitable temperature, they break in the middle to form an exact replica of itself. This phenomenon is called the event of *reproduction* in BFOA. However, due to the occurrence of sudden environmental changes or attacks, the chemotaxis progress may be destroyed, and a group of bacteria may move to some other place or some other mutation may be introduced to the bacteria population. This phenomenon is called the *elimination dispersal event* in BFOA, where all the bacteria in the region are killed or a group is dispersed into a new part of the environment.

References [38, 51–56] provide a comprehensive analysis of the classical BFOA in different optimization problems.

## 4.2. Primary steps of the proposed DBFOA

The proposed DBFOA improves upon the classical BFOA by modifying the principal mechanisms to specifically handle the PV re-phasing problem. The modified versions of the principal mechanisms of the algorithm were named as *D-Chemotaxis*, *D-Reproduction*, and *D-Elimination dispersal*. The following subsections discuss these three principal mechanisms which drive the proposed DBFOA. The mapping of the terms in the PV re-phasing problem and the classical BFOA problem are shown in Table 1.

Table 1: Related terminology

| Variable | Definition in PV rephrasing problem | Definition in classical BFOA |
|---|---|---|
| $N_{pv}$ | The number of grid-connected PV systems in the network | The dimension of the search space |
| $S$ | The number of PV configuration initializers | Total population of the bacterium |
| $N_c$ | The number of D-chemotactic steps | The number of chemotactic steps |
| $N_r$ | The maximum number of random phase changing steps performed | The swimming length |
| $N_{re}$ | The number of D-Reproduction steps | The number of reproduction steps |
| $P_{ed}$ | D-elimination dispersal probability | Elimination dispersal probability |
| $i$ | i-th PV configuration vector | The $i$-th bacterium in the population |
| $j$ | Incremental counter (index) for D-chemotaxis step | Index for the chemotaxis step |
| $k$ | Incremental counter (index) for D-reproduction step | Index for the reproduction step |
| $l$ | Incremental counter (index) for D-elimination dispersal step | Index of the elimination-dispersal event |
| $r$ | Incremental counter (index) for the random phase changing step | Index for swimming step |
| $J(i,k,k,l)$ | The cost of $i$-th PV configuration vector $\boldsymbol{x^i}(j,k,l)$ | The cost at the location of the $i$-th bacterium $\boldsymbol{x^i}(j,k,l)$ |

### 4.2.1. Discrete Chemotaxis (D-Chemotaxis):

The D-Chemotaxis step updates the phase combination of a PV configuration vector such that the new phase combination has a lower cost value compared to its previous phase combination. In other words, the D-Chemotaxis step updates the phase combination in a direction corresponding to a gradient of decreasing cost value.

Here, the present phase combination in $i$-th PV configuration vector is given by $\boldsymbol{x^i}(j,k,l)$ and it's updated version is denoted by $\boldsymbol{x^i}(j+1,k,l)$ where, $j$, $k$, and $l$ are the index for D-Chemotaxis, D-Reproduction, and D-Elimination dispersal, respectively.

The proposed D-Chemotaxis step first identifies the highest unbalance region (${\mathbb{H}_{VU}}^{i}$) in the network corresponding to the phase combination in the $i$-the PV configuration vector, $x^i(j,k,l)$. Here, the highest unbalance region is referred to the busbars within $\mathrm{k_n}$ number of busbars from the busbar with highest unbalance ($\mathrm{n_{VUF_{max}}}$). Once the highest unbalance region ${\mathbb{H}_{VU}}^{i}$ is identified, only the phase combinations of grid-connected PV systems in the highest unbalance region, ${\mathbb{H}_{VU}}^{i}$, are randomly changed to generate the updated phase combination, $x^i(j+1,k,l)$. This reduces the number of possible phase configurations greatly, while mitigating the impact of re-phasing on the overall network. However, the random change in the phases of PV systems in the highest unbalance region ${\mathbb{H}_{VU}}^{i}$ does not guarantee that it finds a phase combination with a lower cost value compared to its present phase combination $x^i(j,k,l)$ at once. Therefore, in such a situation, the random phase changing is repeated until it finds a suitable phase combination with lower-cost value, within a maximum of $N_r$ iterations. If D-Chemotaxis is unable to find a phase combination with lesser cost value within the maximum $N_r$ cycles, then the present phase combination $x^i(j,k,l)$ is retained as its updated phase combination $x^i(j+1,k,l)$ as it is reasonable to assume that we have reached a low-cost point through random changes.

The pseudocode of the D-chemotaxis procedure is given in Algorithm 1 and the main steps are depicted in Figure 5.

**Algorithm 1: D-Chemotaxis**

**Step 1:** Perform a load flow analysis for the phase combination in $i$-th PV configuration vector, $\boldsymbol{x}^{\boldsymbol{i}}(j,k,l)$.

**Step 2:** Evaluate the cost function-$J(i,j,k,l)$ based on the load flow results, and set $J_{last} = J(i,j,k,l)$.

**Step 3:** Identify the busbar with the highest voltage unbalance $n_{VUF_{max}}$ and, then identify the busbars within the radius of $k_n$ busbars from the busbar with highest unbalance $n_{VUF_{max}}$ to form the highest unbalance region, $\mathbb{H}_{VU}$.

**Step 4:** Randomly change the phase combination of PV systems in the highest unbalance region $\mathbb{H}_{VU}$ to find a phase combination with lower-cost value compared to $J_{last}$.

**Step 5:** If a suitable phase combination is identified within $N_r$ steps, then use that phase combination as the updated phase combination, $\boldsymbol{x}^i(j+1,k,l)$.

**Step 6:** Else, $\boldsymbol{x}^{\boldsymbol{i}}(j+1,k,l) = \boldsymbol{x}^i(j,k,l)$.

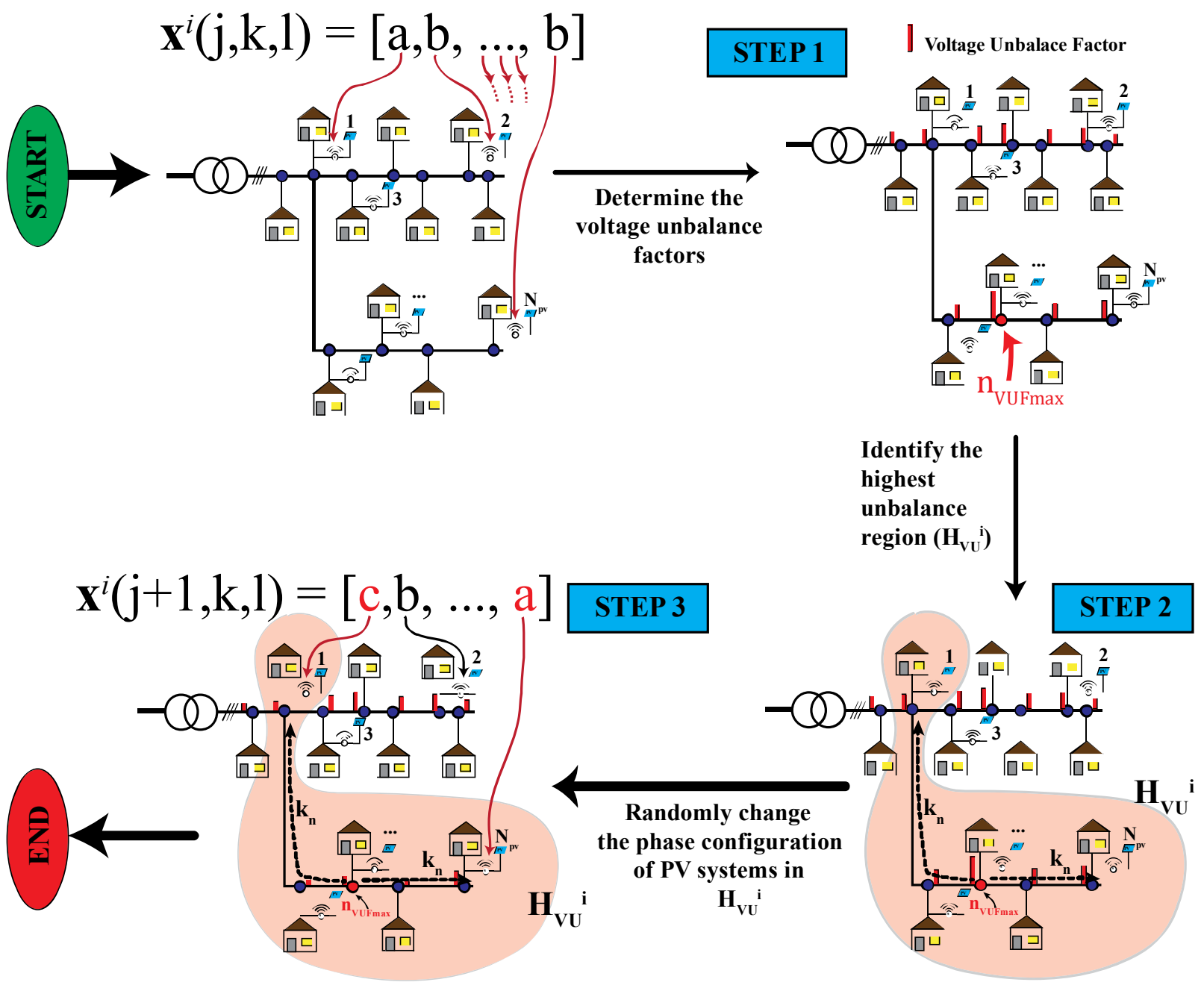

Figure 5: Proposed D-Chemotaxis procedure

#### 4.2.2. Discrete Reproduction (D-Reproduction)

In D-Reproduction, the PV configuration vector having the highest cumulative cost (i.e. worst PV configuration) is eventually replaced by the PV configuration vector with the least cumulative cost (i.e. best PV configuration). The cumulative cost of the $i$-th PV configuration vector ($J_C^i$) for a given $k$ and $l$ was calculated from,

$$J_C^i = \sum_{j=1}^{N_C+1} J(i,j,k,l) \qquad (6)$$

The pseudocode of the D-Reproduction step is given in Algorithm 2.

**Algorithm 2: D-Reproduction**

| | |
|---|---|
| **Step 1:** | Determine the cumulative cost $J_C$ of all the PV configuration vectors for given $k$ and $l$ from Equation **(6)**. |
| **Step 2:** | Replace the phase combination of the PV configuration vector having the highest cumulative cost by the phase combination of the PV configuration vector having the least cumulative cost. |

#### 4.2.3. Discrete Elimination Dispersal (D-Elimination Dispersal)

In D-Elimination Dispersal, some PV configuration vectors are randomly liquidated (eliminated) with a very small probability $P_{ed}$ while the new replacements are randomly initialized over the search space. The D-Elimination Dispersal operator helps PV combinations that are trapped in local minima to escape.

The pseudocode of the D-Elimination Dispersal step is given in Algorithm 3.

**Algorithm 3: D-Elimination Dispersal**

| | |
|---|---|
| **Step 1:** | For all PV configuration vectors (i.e. for $i = 1,2,3,\dots,S$) repeat the following steps to perform D-Elimination dispersal. |
| **Step 2:** | Generate a Random Number between 0 and 1: $RN^i = rand(0,1)$. |
| **Step 3:** | If $RN^i \leq P_{ed}$, Replace the phase combination in $i$-th PV configuration vector by a random phase combination. |
| **Step 4:** | Else, proceed to Step 2 for the next PV configuration vector ($i = i + 1$). |

## 4.3. Initialization of PV configuration vectors

The BFOA is a population-based optimization algorithm. Hence, the quality of the optimal solution and the time to convergence heavily depend on the initial population (in this paper the initial population is also referred to as the set of PV configuration initializers to add more contextual flavor). In most of the situations, the initial population is randomly selected from the solution space. However, it has been noted that random initialization is not an effective way to initialize the PV configuration initializers, especially when more contextual information is available to better optimize the selection of the initial points. Therefore, a novel initialization method was introduced to identify the suitable phase combinations for PV configuration initializers. A performance comparison is added in the results and discussion to highlight the effectiveness of the proposed initialization technique.

The proposed initialization method determines the initial phase combinations for the PV configuration initializers in such a way that those initial phase combinations have a small active power mismatch (see algorithm 4, step 4) at the secondary side of the MV-LV transformer. The suitable phase combinations with minimum active power mismatch were selected from the brute force checking strategy where the active power mismatch for the whole solution space is computed to identify the phase combinations that have smaller active power mismatch. However, for a network that has a large number of grid-connected PV systems, the brute force searching will take a long time to find a set of suitable phase combinations with smaller active power mismatch. Therefore, in order to improve the speed of the initialization, such large networks are partitioned into smaller regions ($R_1, R_2, R_3, \ldots$) and the active power balancing was considered separately for each region (see Figure 9). In other words, a regional minimization is performed to facilitate the global optimization. The proposed initialization process is graphically illustrated in Figure 6 and the pseudocode is given in Algorithm 4.

**Algorithm 4: Initialization of PV configuration initializers**

**Step 1:** Collect active power consumption of loads and the active power generation by PV systems through smart meters.

**Step 2:** Partition the large network into smaller regions ($R_1, R_2, R_3, \ldots, R_r$).

**Step 3:** Identify the solution space (i.e. all possible phase combinations) for each region. Note: For a region having $w$ number of grid-connected PV systems, there are $3^w$ possible phase combinations in the solution space.

**Step 4:** ***Execute in parallel for $R_1, R_2, R_3, \ldots, R_r$:***
**1)** Calculate the active power mismatch for each phase combination in the solution space. The active power mismatch is quantified by the standard deviation of the three-phase active power in that region.
**2)** Identify $k_R(= 4)$ phase combinations having a smallest active power mismatch among the all possible phase combinations.

**Step 5:** Randomly combine the identified phase combinations for each region to form the initial phase combinations for PV configuration initializers.

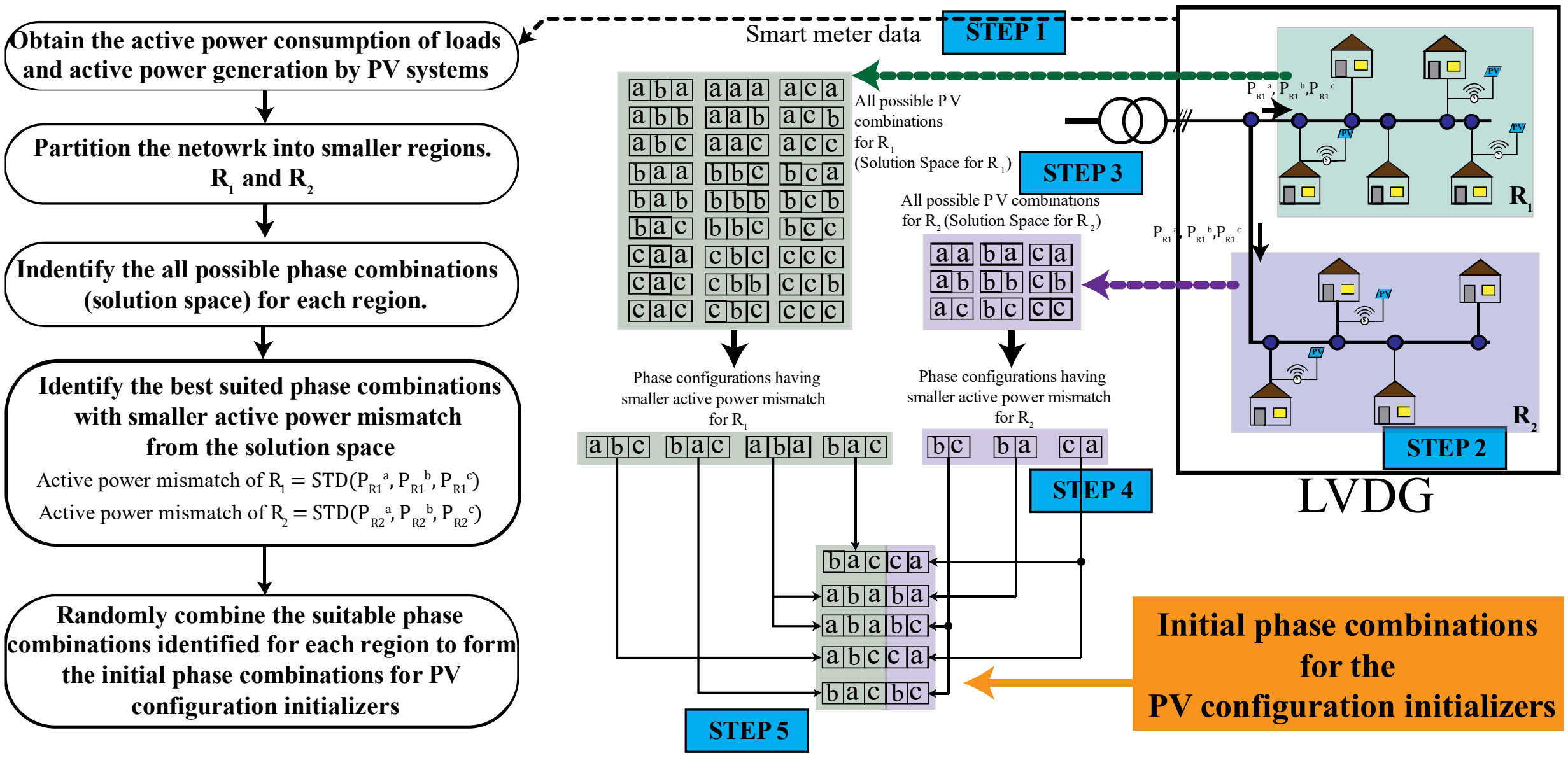

Figure 6: Generation of initial phase combinations for PV configuration initializers based on active power balancing technique for a large distribution network.

## 4.4. Implementation of DBFOA

### 4.4.1. The complete structure of the proposed DBFOA

The complete structure of the proposed DBFOA is shown in Figure 7.

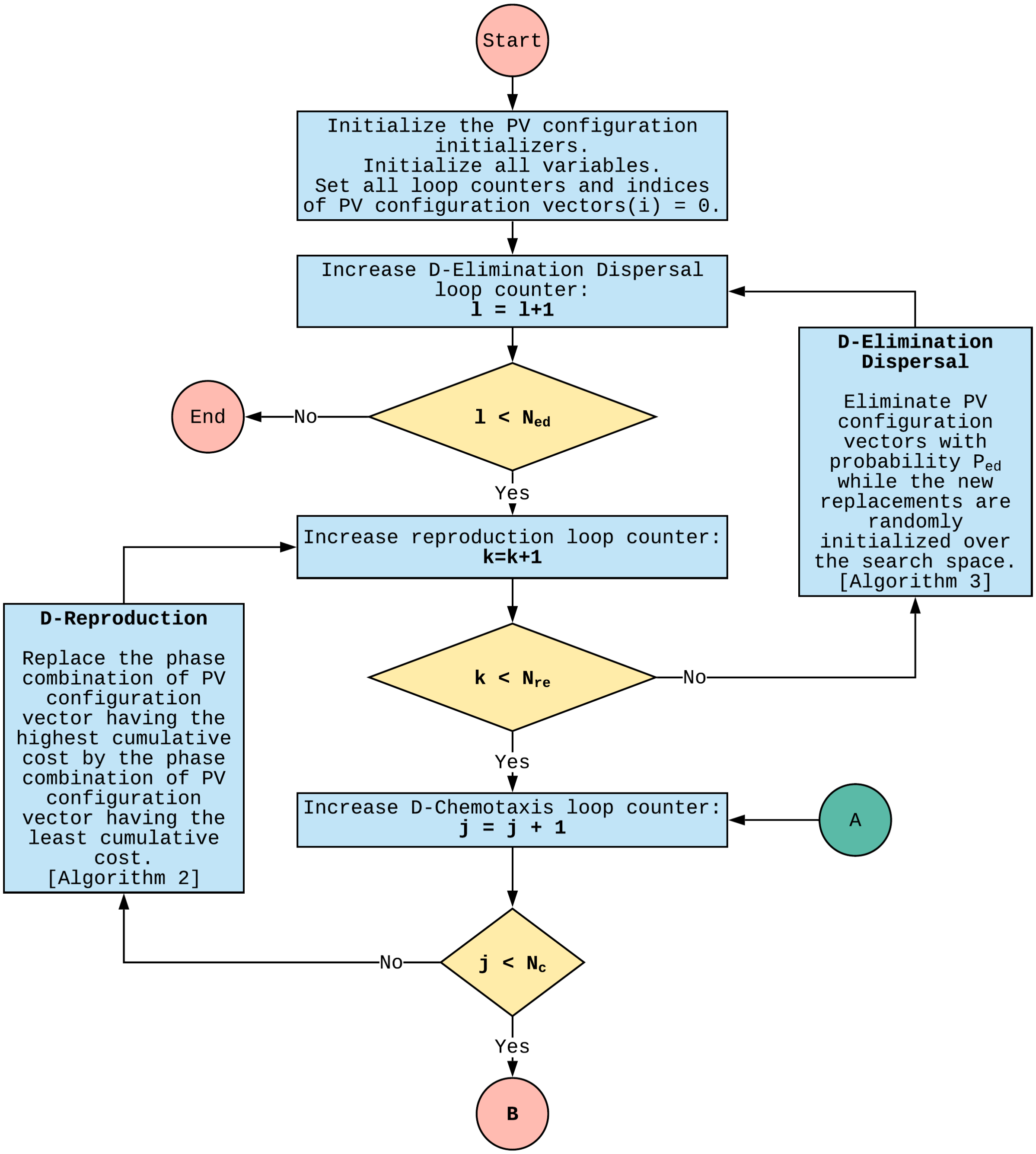

Figure 7: Continued.

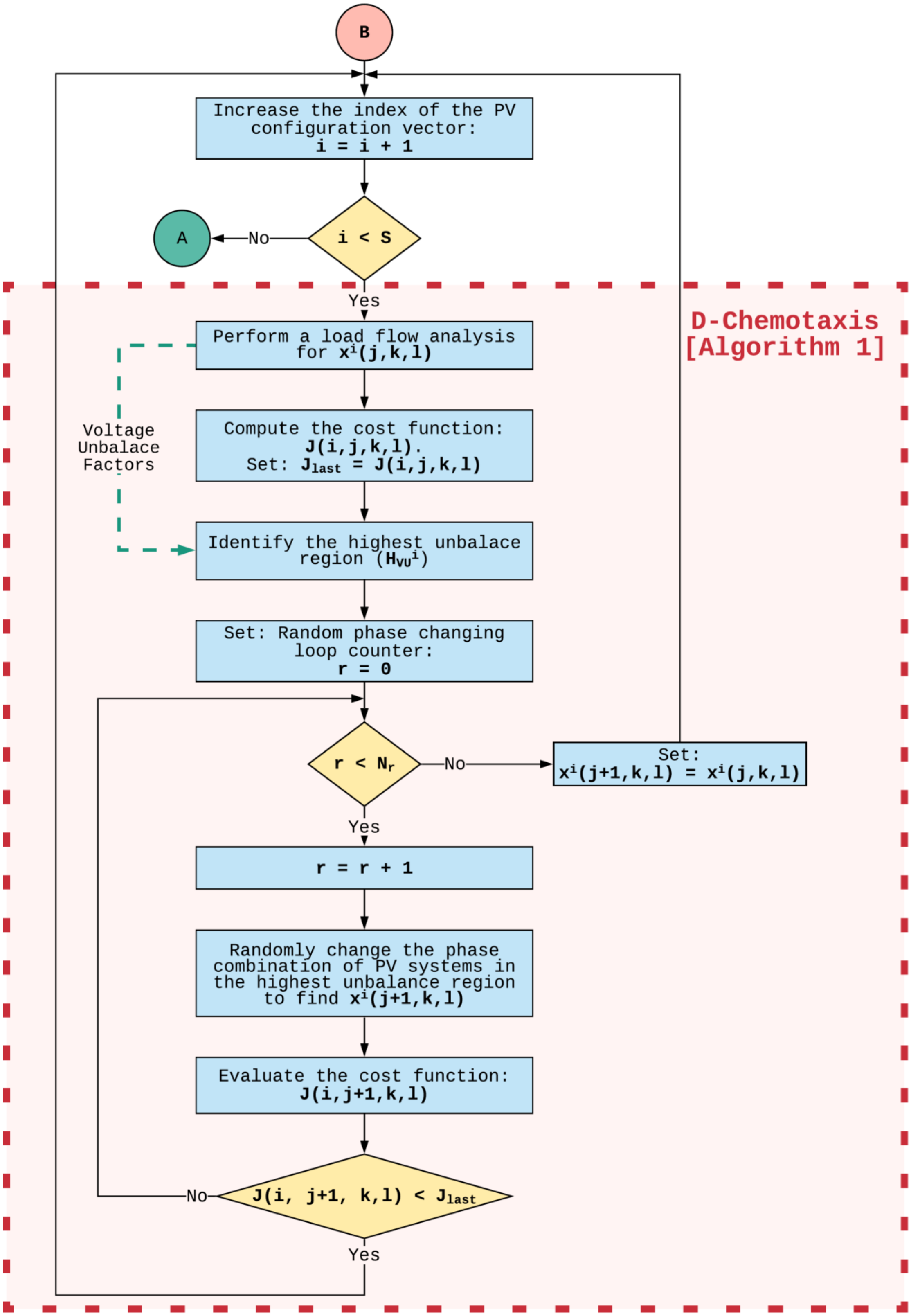

Figure 7: The complete structure of the proposed DBFOA

#### 4.4.2. Pseudocode of DBFOA

The pseudocode of the proposed DBFOA applied to reduce overall unbalance of a network is given in Algorithm 5.

---

**Algorithm 5: The proposed DBFOA**

---

**Step 1:** **Initialize all the PV configuration initializers - $x^i$ (use Algorithm 4).**
**Initialize of the following parameters:**

- $S$: The number of PV configuration initializers.
- $N_c$: The maximum number of D-Chemotaxis is performed.
- $N_s$: The Maximum number of times random phase changing is performed.
- $N_{re}$: The maximum number of times D-Reproduction is performed.
- $P_{ed}$: The probability that each *PV configuration vector*s are eliminated.

**Set all loop counters to zero.**

- Incremental counter for D-Elimination dispersal step $(l) = 0$.
- Incremental counter for D-Reproduction step $(k) = 0$.
- Incremental counter for D-Chemotaxis step $(j) = 0$.
- Index of the PV configuration vectors $(i) = 0$.

**Step 2:** D-Elimination Dispersal loop: $l = l + 1$.

**Step 3:** D-Reproduction loop: $k = k + 1$.

**Step 4:** **D-Chemotaxis loop**, $j = j + 1$. **[Algorithm 1]**

- **A.** **For** $i = 1,2, \ldots, S$, execute a D-Chemotaxis step for $i$-th PV configuration vector as follows.
- **B.** Perform a load flow analysis for the phase combination in $i$-th PV configuration vector - $\boldsymbol{x}^i(j, k, l)$ to obtain three-phase voltages and voltage unbalance factors.
- **C.** Evaluate the cost function $J(i, j, k, l)$.
- **D.** Let $J_{last} = J(i, j, k, l)$ so that the PV configuration vector having a lower cost could be identified.
- **E.** Identify the highest voltage unbalance region $\mathbb{H}_{VU}^{i}$ corresponding to phase combination $\boldsymbol{x}^i(j, k, l)$ from the voltage unbalance factors obtained in **Step 4 B**.
- **F.** Set: random phase changing loop counter to zero, $r = 0$.
- **G.** **While** $r < N_r$,
  - i. Increment the random phase changing loop counter: $r = r + 1$.
  - ii. Randomly change the phase combination of the PV systems in the highest unbalance region to determine the new phase combination of $i$-th PV configuration vector $\boldsymbol{x}^i(j + 1, k, l)$.

iii. Evaluate the cost function $J(i, j+1, k, l)$ corresponding to the phase combination $\boldsymbol{x}^i(j+1, k, l)$.

iv. **If:** $J(i, j+1, k, l) < J_{last}$,
Go to the next PV configuration vector $(i = i + 1)$ (i.e. Go to **Step 4 B.** to process the next PV configuration vector).

v. **Else:**
Go to **Step 4 G.**

**H. End of while**.
Couldn't find a phase combination better than $\boldsymbol{x}^i(j, k, l)$.
Set $\boldsymbol{x}^i(j+1, k, l) = \boldsymbol{x}^i(j, k, l)$.
Go to the next PV configuration vector $(i = i + 1)$ (i.e. Go to **Step 4 B.** to process the next PV configuration vector).

**Step 5:** **If,** $j < N_c$ go to **Step 4** $(j = j + 1)$. In this case, continue D-Chemotaxis.

**Else,** go to **Step 6**.

**Step 6:** **D-Reproduction** [Algorithm 2]:

A. For the given $k$ and $l$, and for each $i = 1,2, \ldots, S$ evaluate the cumulative cost of $i$-th PV configuration vector as follows:

$$J_C^i = \sum_{j=1}^{N_C+1} J(i, j, k, l)$$

B. Replace the phase combination of the PV configuration vector having the highest cumulative cost by the phase combination of the PV configuration vector having the least cumulative cost.

**Step 7:** **If** $k < N_{re}$, go to **Step 3** $(k = k + 1)$. We have not reached the number of specified D-reproduction steps, so we start the next generation of the D-Chemotaxis loop.

**Else,** go to **Step 8**.

**Step 8:** **D-Elimination Dispersal** [Algorithm 3]:

- **For** $i = 1,2, \ldots, S$, eliminate PV configuration vectors with probability $P_{ed}$ while the new replacements are randomly initialized over the search space.

**Step 9:** **If,** $l < N_{ed}$, go to *Step 2* $(l = l + 1)$;

**Else**, End.

### 4.4.3. Parameters of DBFOA

The parameter values used in the proposed DBFOA are given in Table 2. The number of PV configuration initializers ($S$), and the values for $N_c, N_r, N_{re}$, and $N_{ed}$ were selected by considering the convergence speed of the DBFOA and the values used in the previous studies [38, 51, 53]. The elimination-dispersal probability - $P_{ed}$ and the radius of the highest unbalance region - $k_n$ were selected to maximize the convergence speed of DBFOA by executing the algorithm for a possible range of values for $P_{ed}$ and $k_n$.

Table 2: Parameter values used for the proposed DBFOA

| Parameter | Value |
|---|---|
| Number of PV configuration initializers ($S$) | 10 |
| Maximum number of D-Chemotaxis steps ($N_c$) | 5 |
| Maximum number of random phase changing steps ($N_r$) | 5 |
| Maximum number of D-Reproduction steps ($N_{re}$) | 5 |
| Maximum number of D-Elimination steps ($N_{ed}$) | 5 |
| Elimination & dispersal probability ($P_{ed}$) | 0.2 |
| The radius of the highest voltage unbalance region - $\mathbb{H}_{VU}$ ($k_n$) | 3 |
| The maximum limit for phase voltage magnitudes ($V_{max}$) | 1.06 pu |
| The minimum limit for phase voltage magnitudes ($V_{min}$) | 0.94 pu |
| The maximum limit for voltage unbalance factors ($VUF_{max}$) | 1% |
| Number of grid-connected PV systems in the network ($N_{pv}$) | 26 |
| Number of busbars in the network ($N$) | 63 |

## 5. Test network

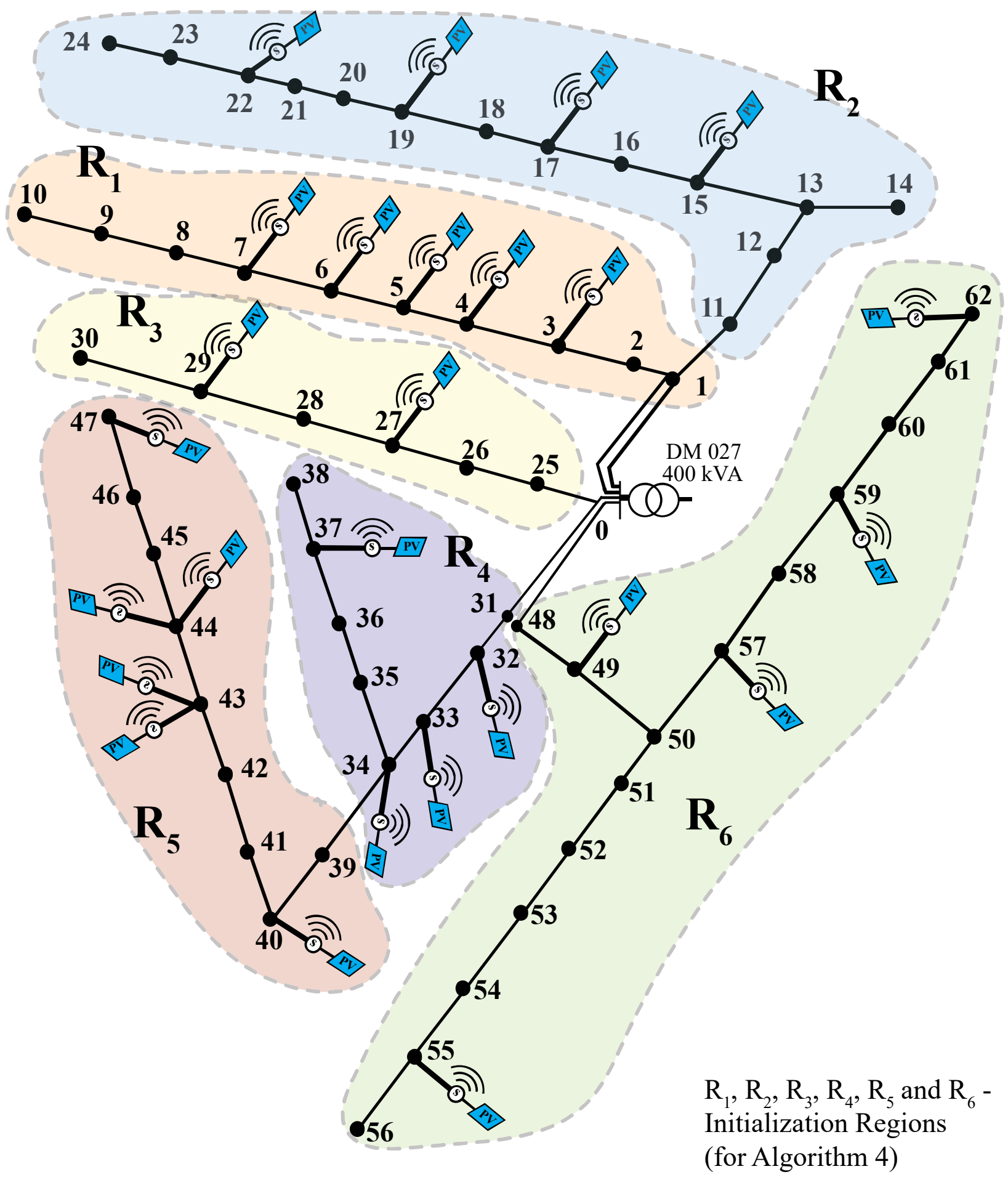

Figure 8: Single line diagram of the test LVDG network (Lotus Grove, Sri Lanka) used for the simulations.

The network topology with 63 busbars as shown in Figure 8 was used for the simulations. The number 0 node is the root node and connected to the secondary side of the MV-LV transformer. The rated capacity of the transformer is 400 kVA and the input/output voltage rating is 11 kV/415 V. The solid lines in Figure 8 represent the three-phase feeders with three-phase or single-phase loads and PV systems connected. The per length impedance matrix of the feeder line is given in Table 3.

Table 3: Impedance matrix ($Z^{abcn}$) per km for the overhead cable (ABC 70) in the network.

| | Phase a | Phase b | Phase c | neutral |
|---|---|---|---|---|
| Phase a | $0.4918 + 0.7888i$ | $0.0486 + 0.6292i$ | $0.0487 + 0.6701i$ | $0.0486 + 0.7000i$ |
| Phase b | $0.0486 + 0.6292i$ | $0.4918 + 0.7888i$ | $0.0487 + 0.6405i$ | $0.0486 + 0.6490i$ |
| Phase c | $0.0487 + 0.6701i$ | $0.0487 + 0.6405i$ | $0.4918 + 0.7888i$ | $0.0487 + 0.7080i$ |
| neutral | $0.0486 + 0.7000i$ | $0.0486 + 0.6490i$ | $0.0487 + 0.7080i$ | $0.6790 + 0.7910i$ |

There are 26 grid-connected single-phase PV systems and 92 single-phase or three-phase loads. The capacity of PV systems, their locations, and their default phase configuration are given in Table 4 and the daily operation curve (hourly generation profile) of PV systems is shown in Figure 9-(a).

Table 4: The details of the grid-connected single-phase PV systems connected to the network

| PV No. | Connected busbar | Initial phase configuration (Fixed phase) | Capacity ($P_{max}$)/ kW |
|---|---|---|---|
| PV1 | **3** | a | 2.40 |
| PV2 | **4** | b | 7.20 |
| PV3 | **5** | a | 4.80 |
| PV4 | **6** | c | 2.40 |
| PV5 | **7** | a | 6.00 |
| PV6 | **15** | a | 5.04 |
| PV7 | **17** | b | 5.04 |
| PV8 | **19** | a | 2.40 |
| PV9 | **22** | c | 8.20 |
| PV10 | **27** | a | 4.80 |
| PV11 | **29** | c | 6.48 |
| PV12 | **32** | a | 6.00 |
| PV13 | **33** | b | 3.96 |
| PV14 | **34** | a | 6.00 |
| PV15 | **37** | b | 3.12 |
| PV16 | **40** | c | 4.20 |
| PV17 | **43** | c | 6.00 |
| PV18 | **43** | b | 8.2 |
| PV19 | **44** | c | 6.36 |
| PV20 | **44** | b | 5.76 |
| PV21 | **47** | a | 8.28 |
| PV22 | **49** | a | 5.16 |
| PV23 | **55** | c | 8.40 |
| PV23 | **57** | b | 2.20 |
| PV25 | **59** | a | 4.80 |
| PV26 | **62** | c | 7.20 |

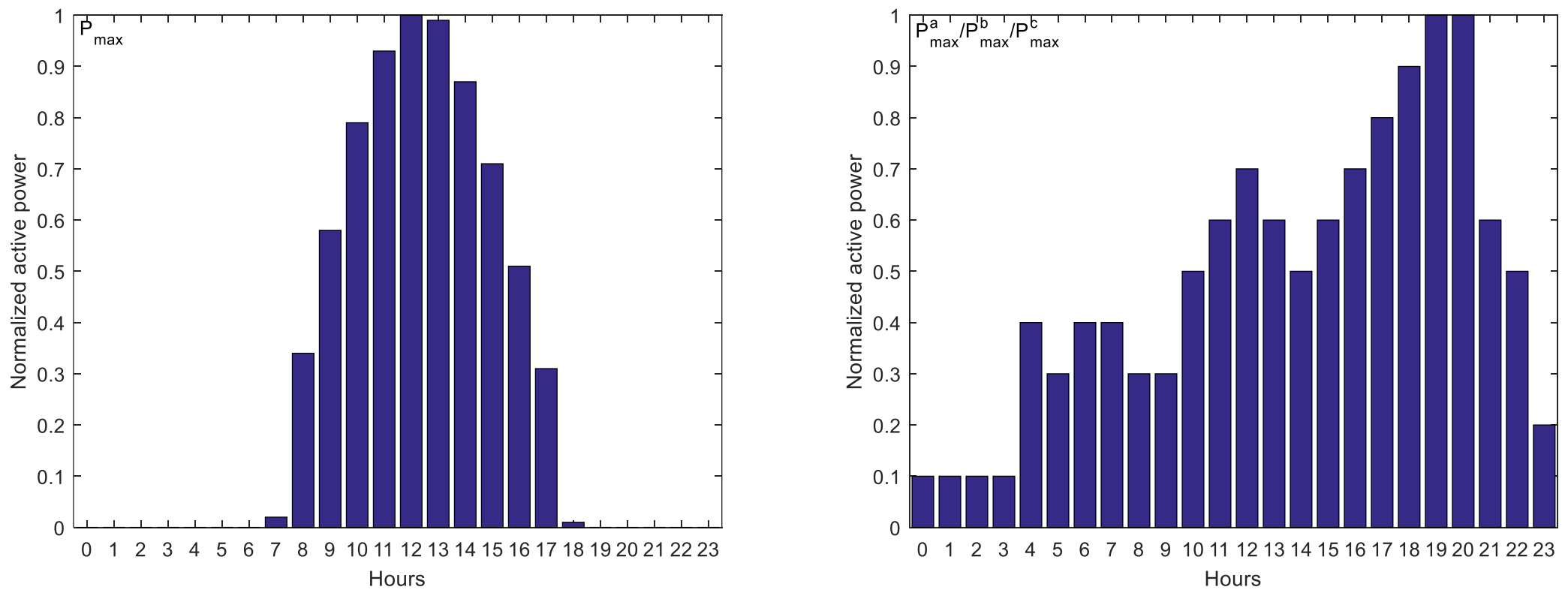

Figure 9: (a) Hourly generation profile of PV systems (b) hourly load profile for single-phase/ three-phase loads.

Moreover, the network shown in Figure 8 consists of 92 single-phase and three-phase loads. The maximum capacity of loads and their power factor values at each busbar are given in Table 5, and the hourly load profile of loads is shown in Figure 9-(b).

Table 5: The capacity of loads and power factor values at each busbar

| Busbar | Active power/ (kW) | | | pf |
|---|---|---|---|---|
| | $P_{max}^{a}$ (kW) | $P_{max}^{b}$ (kW) | $P_{max}^{ac}$ (kW) | |
| **1** | 2.19 | 0.55 | 1.85 | 0.981 |
| **1** | 1.28 | 0.46 | 0.55 | 0.991 |
| **1** | 2.43 | 0.28 | 1.88 | 0.913 |
| **2** | 2.33 | 0.26 | 1.50 | 0.991 |
| **3** | 0.19 | 0.45 | 0.34 | 0.963 |
| **3** | 0.12 | 0.69 | 0.68 | 0.910 |
| **4** | 0.33 | 0.58 | 0.17 | 0.928 |
| **4** | 0.48 | 0.52 | 1.08 | 0.955 |
| **5** | 0.54 | 1.36 | 0.69 | 0.996 |
| **5** | 1.43 | 1.43 | 1.33 | 0.996 |
| **6** | 1.53 | 2.22 | 0.64 | 0.916 |
| **6** | 1.04 | 0.78 | 0.27 | 0.997 |
| **7** | 0.24 | 0.25 | 0.29 | 0.996 |
| **7** | 0.56 | 1.24 | 0.79 | 0.949 |
| **8** | 0.54 | 1.09 | 2.46 | 0.980 |
| **8** | 1.76 | 0.42 | 0.90 | 0.914 |
| **9** | 1.03 | 2.79 | 0.86 | 0.942 |
| **9** | 1.18 | 0.51 | 0.99 | 0.992 |
| **10** | 0.91 | 4.05 | 0.06 | 0.979 |
| **13** | 1.39 | 1.45 | 0.74 | 0.996 |
| **14** | 0.16 | 0.38 | 0.15 | 0.966 |
| **15** | 4.93 | 1.57 | 0.14 | 0.904 |
| **15** | 1.39 | 2.17 | 0.32 | 0.985 |
| **16** | 0.96 | 0.90 | 0.53 | 0.993 |
| **16** | 1.44 | 2.95 | 0.50 | 0.968 |
| **17** | 0.77 | 0.69 | 0.72 | 0.976 |
| **17** | 0.92 | 1.05 | 0.82 | 0.974 |
| **18** | 0.21 | 0.74 | 1.13 | 0.939 |
| **18** | 1.03 | 1.18 | 1.88 | 0.966 |
| **19** | 0.34 | 2.01 | 1.63 | 0.917 |
| **19** | 1.51 | 1.04 | 1.44 | 0.971 |
| **20** | 1.14 | 1.03 | 0.21 | 0.903 |
| **21** | 0.73 | 0.58 | 0.97 | 0.928 |
| **22** | 1.37 | 1.89 | 0.52 | 0.905 |
| **22** | 1.32 | 1.94 | 1.02 | 0.910 |
| **23** | 1.28 | 1.45 | 1.56 | 0.982 |
| **23** | 0.40 | 0.27 | 0.21 | 0.969 |
| **24** | 0.26 | 0.07 | 0.26 | 0.932 |
| **25** | 0.90 | 1.99 | 0.69 | 0.995 |
| **26** | 2.04 | 1.80 | 0.84 | 0.903 |
| **27** | 4.62 | 0.12 | 0.30 | 0.944 |
| **28** | 0.52 | 2.18 | 1.98 | 0.938 |
| **29** | 0.27 | 0.59 | 0.42 | 0.977 |
| **30** | 1.35 | 2.42 | 0.72 | 0.980 |
| **35** | 2.25 | 0.92 | 0.82 | 0.919 |

| Busbar | Active power/ (kW) | | | pf |
|---|---|---|---|---|
| | $P_{max}^{a}$ (kW) | $P_{max}^{b}$ (kW) | $P_{max}^{c}$ (kW) | |
| **35** | 0.85 | 0.88 | 0.75 | 0.949 |
| **36** | 0.67 | 1.45 | 1.27 | 0.945 |
| **37** | 0.79 | 0.66 | 1.24 | 0.965 |
| **37** | 1.91 | 1.14 | 1.63 | 0.971 |
| **38** | 1.97 | 1.55 | 1.27 | 0.975 |
| **41** | 1.72 | 1.05 | 2.01 | 0.928 |
| **41** | 1.35 | 1.15 | 1.39 | 0.968 |
| **42** | 0.47 | 0.60 | 0.32 | 0.966 |
| **42** | 1.27 | 1.50 | 1.11 | 0.916 |
| **43** | 1.97 | 0.30 | 2.12 | 0.912 |
| **43** | 1.27 | 1.04 | 0.87 | 0.950 |
| **44** | 0.98 | 0.25 | 1.86 | 0.996 |
| **44** | 1.72 | 0.65 | 0.41 | 0.934 |
| **45** | 1.06 | 1.71 | 1.81 | 0.959 |
| **45** | 0.70 | 0.76 | 1.02 | 0.922 |
| **45** | 0.41 | 2.49 | 1.99 | 0.975 |
| **45** | 0.69 | 1.43 | 1.07 | 0.926 |
| **46** | 0.46 | 0.15 | 0.18 | 0.951 |
| **46** | 2.11 | 1.02 | 1.85 | 0.970 |
| **46** | 1.01 | 1.88 | 1.50 | 0.989 |
| **47** | 1.22 | 1.05 | 1.22 | 0.996 |
| **47** | 2.59 | 1.19 | 1.10 | 0.955 |
| **47** | 1.39 | 2.61 | 0.29 | 0.914 |
| **50** | 2.03 | 0.89 | 1.17 | 0.915 |
| **51** | 2.05 | 0.87 | 1.86 | 0.926 |
| **51** | 0.48 | 0.17 | 0.54 | 0.984 |
| **52** | 0.42 | 0.43 | 0.24 | 0.925 |
| **52** | 1.20 | 1.57 | 1.52 | 0.981 |
| **53** | 0.59 | 0.29 | 0.81 | 0.924 |
| **53** | 1.76 | 2.61 | 0.42 | 0.993 |
| **54** | 0.78 | 1.12 | 0.89 | 0.935 |
| **54** | 3.42 | 1.45 | 0.18 | 0.920 |
| **55** | 0.16 | 0.04 | 0.19 | 0.925 |
| **55** | 0.13 | 0.09 | 0.06 | 0.962 |
| **56** | 2.36 | 1.45 | 0.97 | 0.947 |
| **57** | 0.76 | 1.03 | 1.79 | 0.935 |
| **57** | 1.56 | 0.83 | 0.99 | 0.983 |
| **58** | 1.69 | 0.33 | 1.56 | 0.959 |
| **58** | 1.32 | 1.37 | 1.60 | 0.955 |
| **59** | 0.98 | 1.85 | 3.82 | 0.992 |
| **59** | 0.95 | 0.68 | 0.75 | 0.929 |
| **60** | 0.99 | 1.35 | 1.45 | 0.976 |
| **60** | 0.74 | 1.04 | 0.81 | 0.975 |
| **61** | 0.98 | 1.85 | 3.82 | 0.938 |
| **62** | 0.95 | 0.68 | 0.75 | 0.957 |

# 6. Results and Discussion

## 6.1. Convergence characteristics of the proposed DBFOA

### 6.1.1. The effect of D-Chemotaxis

This section demonstrates the effectiveness of the D-Chemotaxis procedure utilized in DBFOA. The proposed D-Chemotaxis is specially designed for the PV re-phasing problem as opposed to merely directly adopting it from classical chemotaxis. The proposed D-Chemotaxis identifies the region around the busbar with the highest unbalance ($\mathbb{H}_{VU}$) and then, only the phase combination of grid-connected PV systems in that region are randomly changed to find the optimal solution iteratively. However, in classical chemotaxis, a fixed number of PV systems are randomly selected from the network and then, the phase configurations of those PV systems are randomly changed. This random selection of PV systems in classical chemotaxis may result in the formation of much higher unbalance levels in the network, thus, ultimately resulting in slower convergence.

Since the increase in voltage unbalance levels of a particular region of a network is mainly due to the mismatch of active and reactive power levels in the same region, the proposed D-chemotaxis step randomly changes the phase configuration of PV systems in the highest unbalance region iteratively. This could lead to a dramatic increase in the convergence speed of the DBFOA as shown in Figure 10. According to the results, the proposed D-chemotaxis step in the DBFOA resulted in faster convergence of the algorithm when compared to the classical chemotaxis under the same conditions (same initial population, same values for parameters, etc.). Also, the proposed chemotaxis step causes the optimal solution to settle in a place with lower cost value as opposed to classical chemotaxis where final settling cost is much higher as Figure 10 depicts.

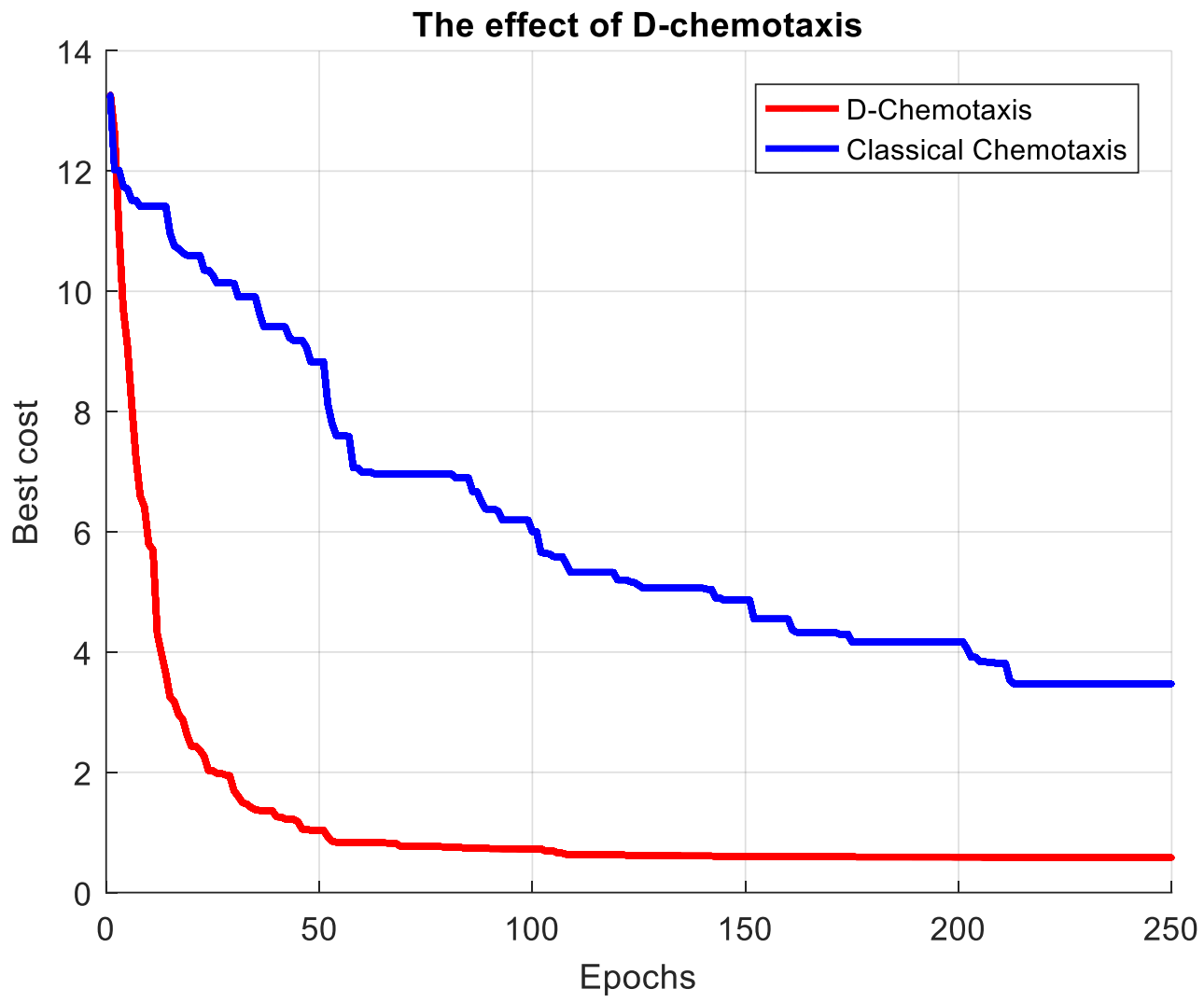

Figure 10: The effect of D-chemotaxis on the convergence of DBFOA.

### 6.1.2. Effect of the initialization of the PV configuration initializers

As described in section 4.3, an active power balancing approach was introduced to find initial phase combinations for the PV configuration initializers as opposed to the random initialization of the classical technique. Figure 11 depicts the convergence properties of the DBFOA under two cases: (1). The PV configuration initializers were initialized based on the active power balancing technique and (2). The PV configuration initializers were randomly initialized. According to the results, the starting cost of DBFOA under the proposed power balancing initialization is approximately 83% lower than the random initialization. This implies that the active power balancing technique was able to generate phase combinations that are closer to the optimal phase combination. Ultimately, the DBFOA with the proposed initialization technique converged to the optimal solution with fewer iterations compared to the random initialization.

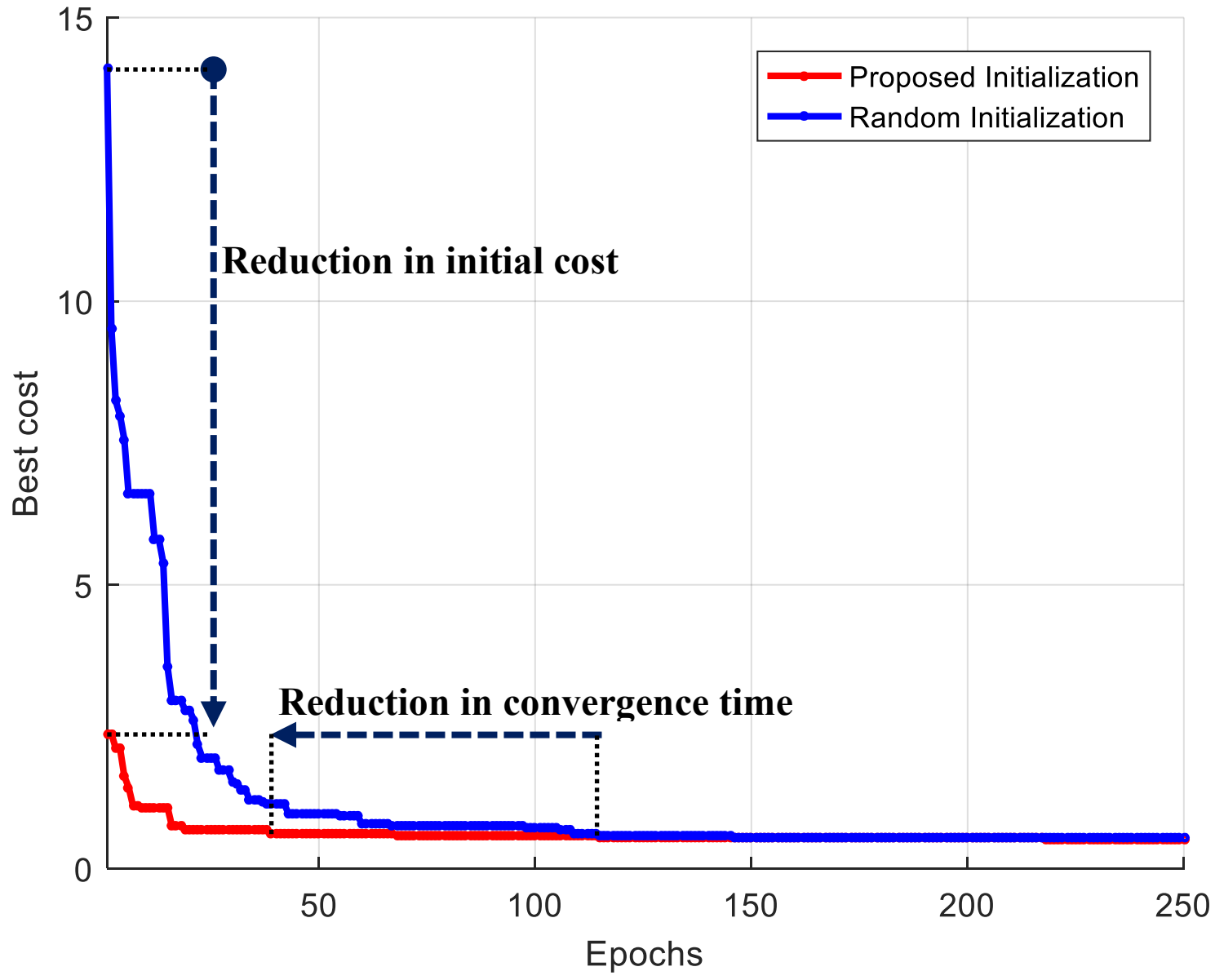

Figure 11: Convergence of the DBFOA for proposed and random initialization methods.

## 6.2. Effects of PV re-phasing

This section demonstrates the effect of PV re-phasing on voltage unbalance and phase voltage magnitudes of residential distribution grid with high penetration of solar power. The proposed PV re-phasing algorithm was implemented on the real distribution network shown in Figure 8. The variations of loads and PV power generation throughout the day were considered in the simulations by utilizing hourly load and PV generation profiles shown in Figure 9.

The goal of PV re-phasing is to reduce overall voltage unbalance ($\overline{\mathrm{VUF}}$) of the network by dynamically changing the phase configuration of rooftop solar systems in the network. Figure 12 and 13 clearly illustrate the effect of PV re-phasing on the overall voltage unbalance and phase voltages, respectively. According to Figure 12, significantly high voltage unbalances are observed during the daytime when PV systems have a fixed phase configuration. Whereas the dynamic PV re-phasing significantly reduces the overall voltage unbalance of the network (mean unbalance is below 1 %) especially during the period from 8 am to 5 pm, where high PV penetration is present. In addition, Table 6 provides the optimal phase configuration of PV systems that are determined from the proposed DBFOA from 6 am to 7 pm. The colored cells in Table 6 belong to the phases of PV systems that are not changed in the subsequent hour. Since not each rooftop PV system is subjected to PV re-phasing at each hour, SCADA needs to send the PV re-phasing commands only for the rooftop solar systems that are required to re-phase in the next PV-rephasing operation. Moreover, the proposed algorithm can be improved to minimize the number of PV re-phasing operations that each rooftop PV system undergoes throughout the day by modifying the cost function of the optimization algorithm. Finally, these results confirm that the proposed re-phasing strategy is successful in reducing the voltage unbalance levels of domestic distribution grids that have many grid-connected rooftop PV systems, while simultaneously maintaining the phase voltages within their acceptable limits.

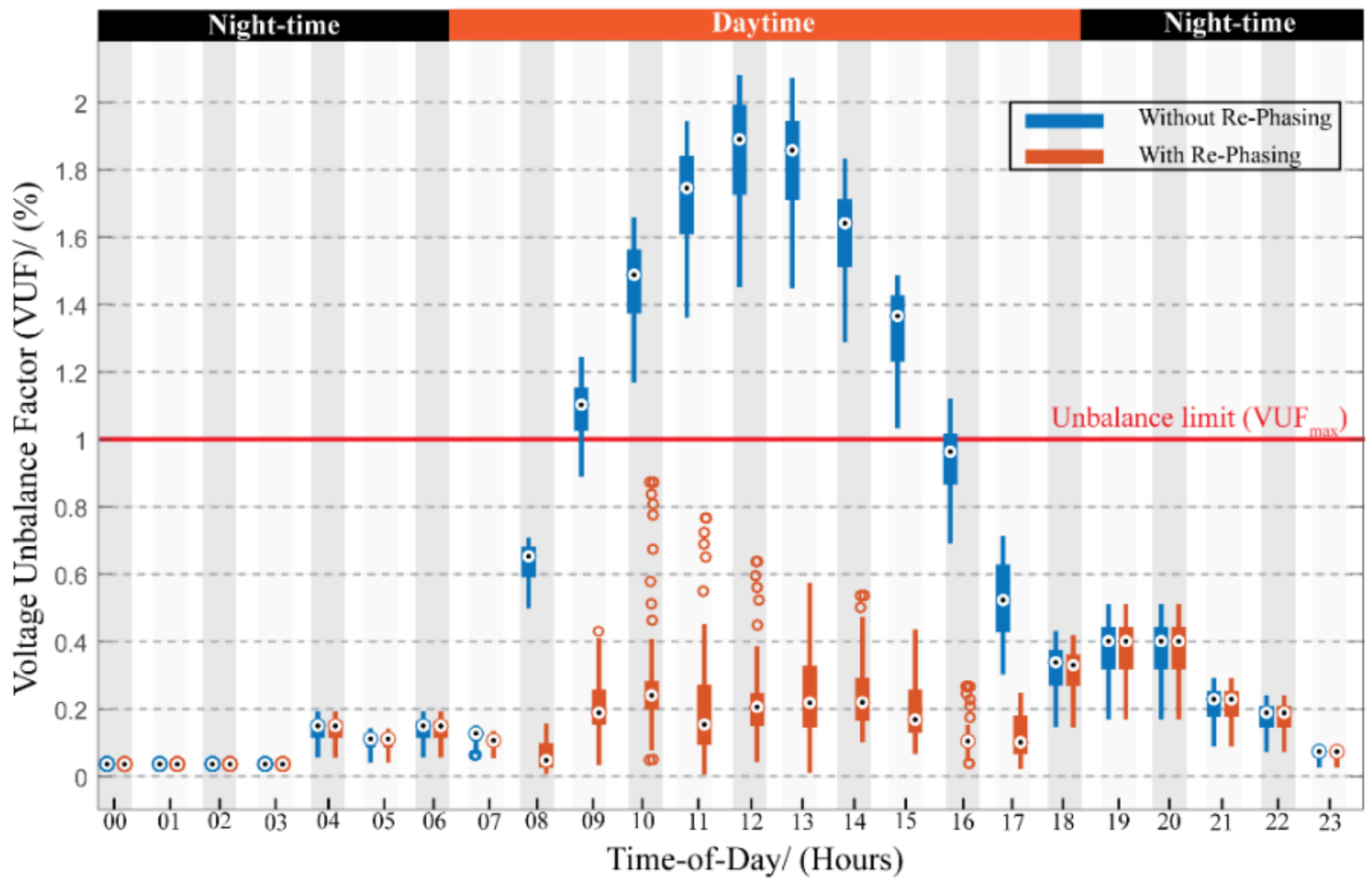

Figure 12: Distribution of voltage unbalance values of the network throughout the day for ‘with’ and ‘without’ PV re-phasing.

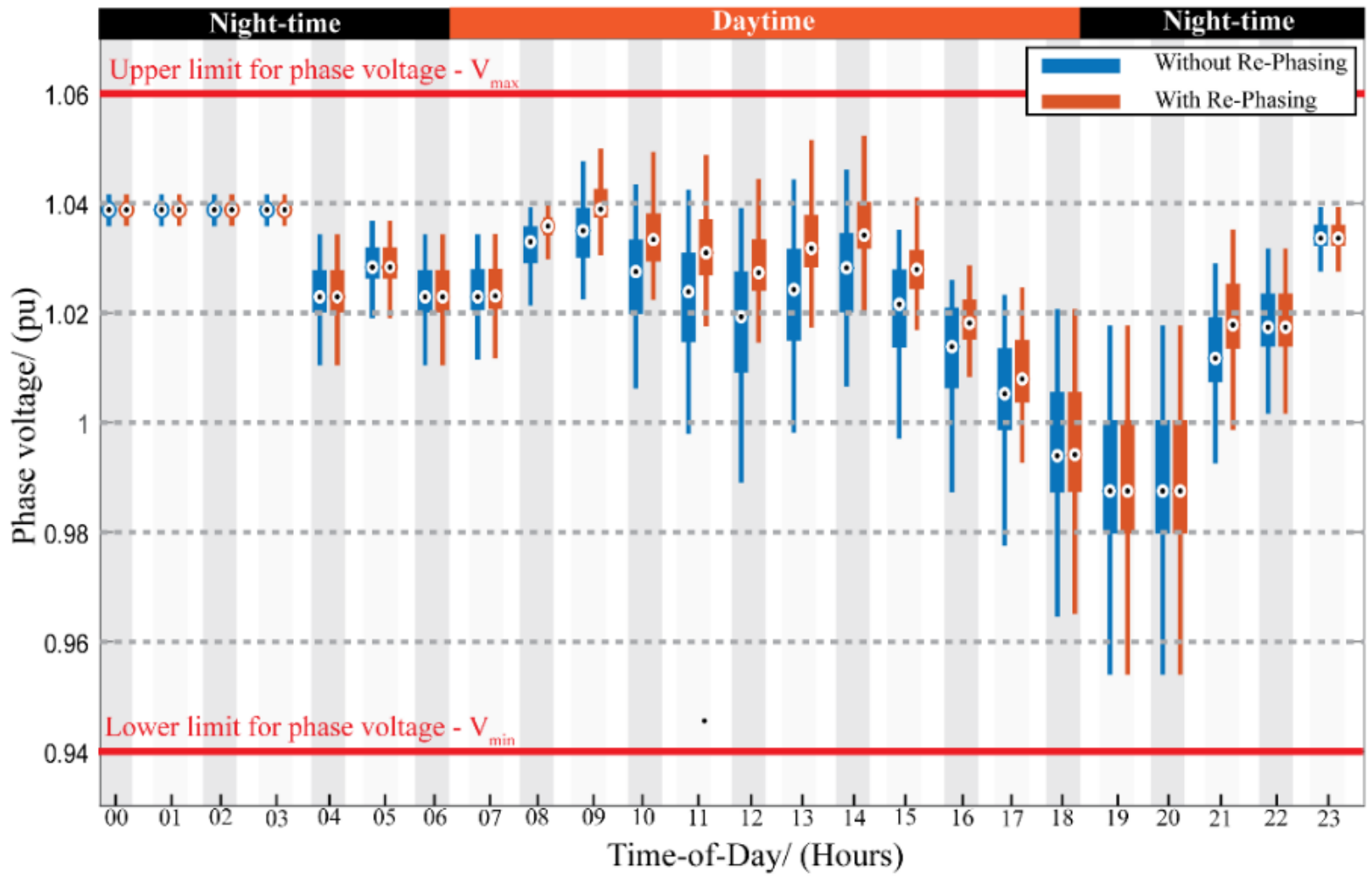

Figure 13: Distribution of phase voltage magnitudes of the network throughout the day for ‘with’ and ‘without’ PV re-phasing.

Table 6: Hourly phase configurations of rooftop PV systems determined by the proposed DBFOA. The colored cells in the table represent the phase configuration of rooftop solar systems that are not changed in the subsequent hour.

| Time | Hourly phase configuration of rooftop PV systems (a = Phase a, b = Phase b, and c= Phase c) | | | | | | | | | | | | | | | | | | | | | | | | |
|---|---|---|---|---|---|---|---|---|---|---|---|---|---|---|---|---|---|---|---|---|---|---|---|---|---|
| | 1 | 2 | 3 | 4 | 5 | 6 | 7 | 8 | 9 | 10 | 11 | 12 | 13 | 14 | 15 | 16 | 17 | 18 | 19 | 20 | 21 | 22 | 23 | 24 | 25 | 26 |
| 6 - 7 am | b ↓ | b ↓ | a ↓ | a ↓ | c ↓ | b ↓ | c ↓ | c ↓ | b ↓ | b ↓ | a ↓ | b ↓ | c ↓ | a ↓ | a ↓ | a ↓ | a ↓ | a ↓ | a ↓ | a ↓ | a ↓ | b ↓ | a ↓ | b ↓ | b ↓ | a ↓ |
| 7 - 8 am | a ↓ | a ↓ | a ↓ | b ↓ | a ↓ | a ↓ | b ↓ | a ↓ | a ↓ | c ↓ | a ↓ | c ↓ | a ↓ | a ↓ | c ↓ | b ↓ | a ↓ | a ↓ | a ↓ | a ↓ | a ↓ | a ↓ | b ↓ | b ↓ | a ↓ | a ↓ |
| 8 - 9 am | a ↓ | a ↓ | a ↓ | a ↓ | a ↓ | a ↓ | a ↓ | b ↓ | b ↓ | a ↓ | a ↓ | a ↓ | b ↓ | b ↓ | a ↓ | a ↓ | a ↓ | a ↓ | a ↓ | a ↓ | b ↓ | b ↓ | b ↓ | a ↓ | a ↓ | a ↓ |
| 9 - 10 am | a ↓ | c ↓ | a ↓ | b ↓ | b ↓ | a ↓ | a ↓ | c ↓ | b ↓ | c ↓ | c ↓ | b ↓ | b ↓ | a ↓ | b ↓ | c ↓ | a ↓ | a ↓ | a ↓ | a ↓ | a ↓ | a ↓ | a ↓ | a ↓ | b ↓ | a ↓ |
| 10 - 11 am | c ↓ | a ↓ | c ↓ | a ↓ | a ↓ | c ↓ | a ↓ | a ↓ | b ↓ | a ↓ | b ↓ | c ↓ | a ↓ | a ↓ | c ↓ | b ↓ | b ↓ | a ↓ | b ↓ | c ↓ | b ↓ | a ↓ | a ↓ | a ↓ | a ↓ | b ↓ |
| 11 - 12 am | a ↓ | a ↓ | a ↓ | a ↓ | b ↓ | a ↓ | a ↓ | a ↓ | a ↓ | c ↓ | c ↓ | b ↓ | c ↓ | b ↓ | b ↓ | c ↓ | a ↓ | b ↓ | a ↓ | a ↓ | a ↓ | a ↓ | a ↓ | a ↓ | a ↓ | a ↓ |
| 12 - 1 pm | b ↓ | a ↓ | a ↓ | a ↓ | a ↓ | a ↓ | a ↓ | c ↓ | c ↓ | a ↓ | b ↓ | a ↓ | a ↓ | a ↓ | b ↓ | a ↓ | a ↓ | a ↓ | a ↓ | b ↓ | a ↓ | b ↓ | a ↓ | a ↓ | b ↓ | b ↓ |
| 1 - 2 pm | a ↓ | b ↓ | a ↓ | a ↓ | a ↓ | a ↓ | a ↓ | a ↓ | b ↓ | b ↓ | c ↓ | b ↓ | b ↓ | b ↓ | a ↓ | b ↓ | a ↓ | a ↓ | b ↓ | a ↓ | c ↓ | b ↓ | a ↓ | a ↓ | a ↓ | a ↓ |
| 2 - 3 pm | a ↓ | a ↓ | b ↓ | a ↓ | a ↓ | a ↓ | a ↓ | a ↓ | a ↓ | c ↓ | a ↓ | c ↓ | a ↓ | a ↓ | c ↓ | a ↓ | a ↓ | a ↓ | a ↓ | a ↓ | b ↓ | a ↓ | c ↓ | b ↓ | a ↓ | a ↓ |
| 3 - 4 pm | a ↓ | a ↓ | a ↓ | a ↓ | a ↓ | a ↓ | a ↓ | b ↓ | b ↓ | b ↓ | a ↓ | b ↓ | b ↓ | b ↓ | b ↓ | b ↓ | a ↓ | a ↓ | a ↓ | a ↓ | b ↓ | b ↓ | a ↓ | a ↓ | a ↓ | a ↓ |
| 4 - 5 pm | a ↓ | a ↓ | a ↓ | a ↓ | a ↓ | a ↓ | a ↓ | c ↓ | b ↓ | b ↓ | c ↓ | b ↓ | b ↓ | b ↓ | a ↓ | c ↓ | a ↓ | a ↓ | a ↓ | a ↓ | b ↓ | b ↓ | a ↓ | a ↓ | a ↓ | a ↓ |
| 5 - 6 pm | a ↓ | a ↓ | a ↓ | a ↓ | b ↓ | b ↓ | a ↓ | a ↓ | a ↓ | b ↓ | a ↓ | a ↓ | c ↓ | c ↓ | b ↓ | b ↓ | a ↓ | b ↓ | a ↓ | a ↓ | b ↓ | a ↓ | a ↓ | a ↓ | a ↓ | a ↓ |
| 6 - 7 pm | a ↓ | a ↓ | a ↓ | a ↓ | a ↓ | a ↓ | a ↓ | a ↓ | b ↓ | a ↓ | b ↓ | b ↓ | a ↓ | b ↓ | a ↓ | a ↓ | a ↓ | a ↓ | a ↓ | a ↓ | a ↓ | b ↓ | a ↓ | a ↓ | a ↓ | a ↓ |
| 7 - 8 pm | a ↓ | b ↓ | a ↓ | b ↓ | a ↓ | a ↓ | b ↓ | b ↓ | c ↓ | b ↓ | b ↓ | b ↓ | c ↓ | c ↓ | c ↓ | c ↓ | a ↓ | a ↓ | b ↓ | a ↓ | c ↓ | c ↓ | a ↓ | a ↓ | b ↓ | b ↓ |

Since the proposed PV re-phasing technique can maintain the voltage imbalances of the network well below the 1% threshold during the daytime while keeping the phase voltages within their acceptable voltage range, utility providers can allow additional rooftop solar systems into the network. In order to get a clear idea about the amount of additional renewable energy capacity facilitated by the PV re-phasing operation, simulations were performed by adding new rooftop solar systems (on top of existing 140.4 kW of solar PV as specified by Table 4) to the existing network. For each addition of a new rooftop solar system, 20 Monte-Carlo simulations were performed by randomly changing its connection point in the LV network to ensure an unbiased and fair simulation. For this study, the capacity of each new rooftop solar system to be connected to the existing LV network is considered to be 5.4 kW that corresponds to the average capacity of a rooftop PV system in the existing network. The maximum voltage unbalances and maximum phase voltage values of the network recorded for "with" and "without" re-phasing are depicted in Figure 14.

According to Figure 14-(c), the maximum voltage unbalance of the network is well below the 1% threshold for the proposed PV re-phasing technique even under the integration of new rooftop solar systems up to a total capacity of 302.4 kW. In contrast, as can be seen from Figure 14-(a), the maximum voltage unbalance values of the network exceed the 1% threshold line for the fixed phase configuration. However, as depicted in Figure 14 – (b) and (d), the maximum voltage of the network gradually increases with the addition of new rooftop solar systems to the network. Due to this reason, rooftop solar systems with a total capacity of 248.4 kW can be safely integrated into the LV network without violating the statutory limits of both voltage unbalance and phase voltage magnitudes. This is about a 77% increase in the rooftop solar capacity of the network compared to the originally installed solar capacity (140.4 kW). Therefore, it is apparent that the proposed PV re-phasing strategy can completely overcome the voltage unbalance issue due to the installation of distributed energy sources in the LV and facilitate to install additional rooftop solar systems into the network.

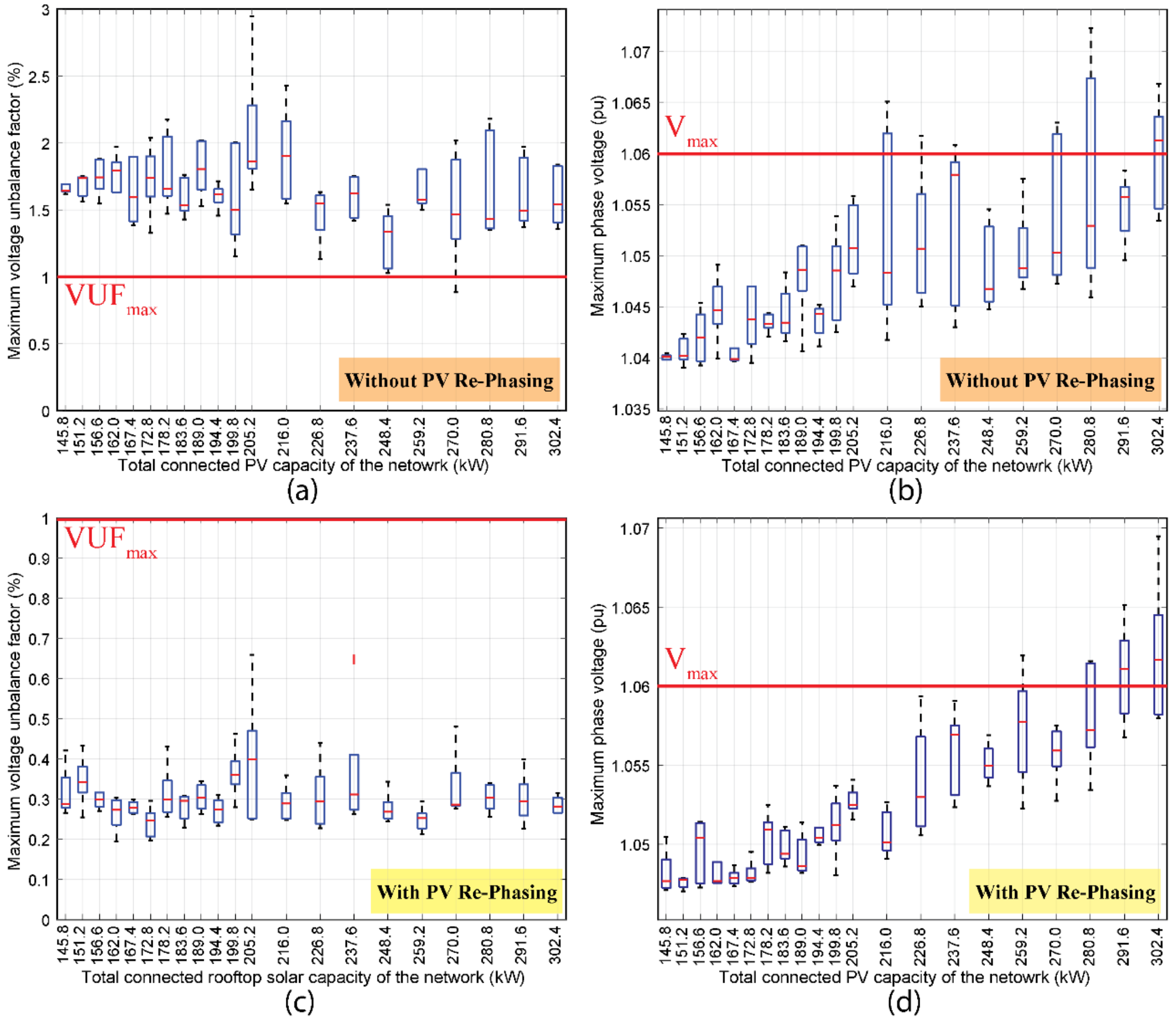

Figure 14: Variation of (a) maximum voltage unbalance and (b) maximum voltage unbalance with the total connected PV capacity of the LVDG network for "without" re-phasing. Variation of (c) maximum voltage unbalance and (d) phase voltage magnitude with the total connected rooftop solar capacity for "with" re-phasing.

## 6.3. Comparison of different optimization techniques

In order to predict the superiority of the proposed DBFOA, the convergence characteristics of the proposed DBFOA for the test system is compared with three other widely used optimization algorithms in power systems, namely, Discrete Genetic Algorithm (DGA), Shuffled Frog-Leaping Algorithm (SFLA) and Heuristic Search (HS), and the results are shown in Figure 15. The algorithms were written on Matlab® (version: R2016a) - Open DSS (version: 8.4.1.1) co-simulation environment and executed on a processor with Intel Core i7-7700HQ with 32 GB RAM running at 3.4 GHz. From the figure, it is clear that the DBFOA only takes 38 iterations to converge to the best solution. In addition to that, DBFOA shows a stable and quick convergence with a global searching capability to find the optimal phase configuration. Thereby, ensuring that the LVDG maintains strict power quality standards, even under heavy PV penetration in a near-real-time fashion.

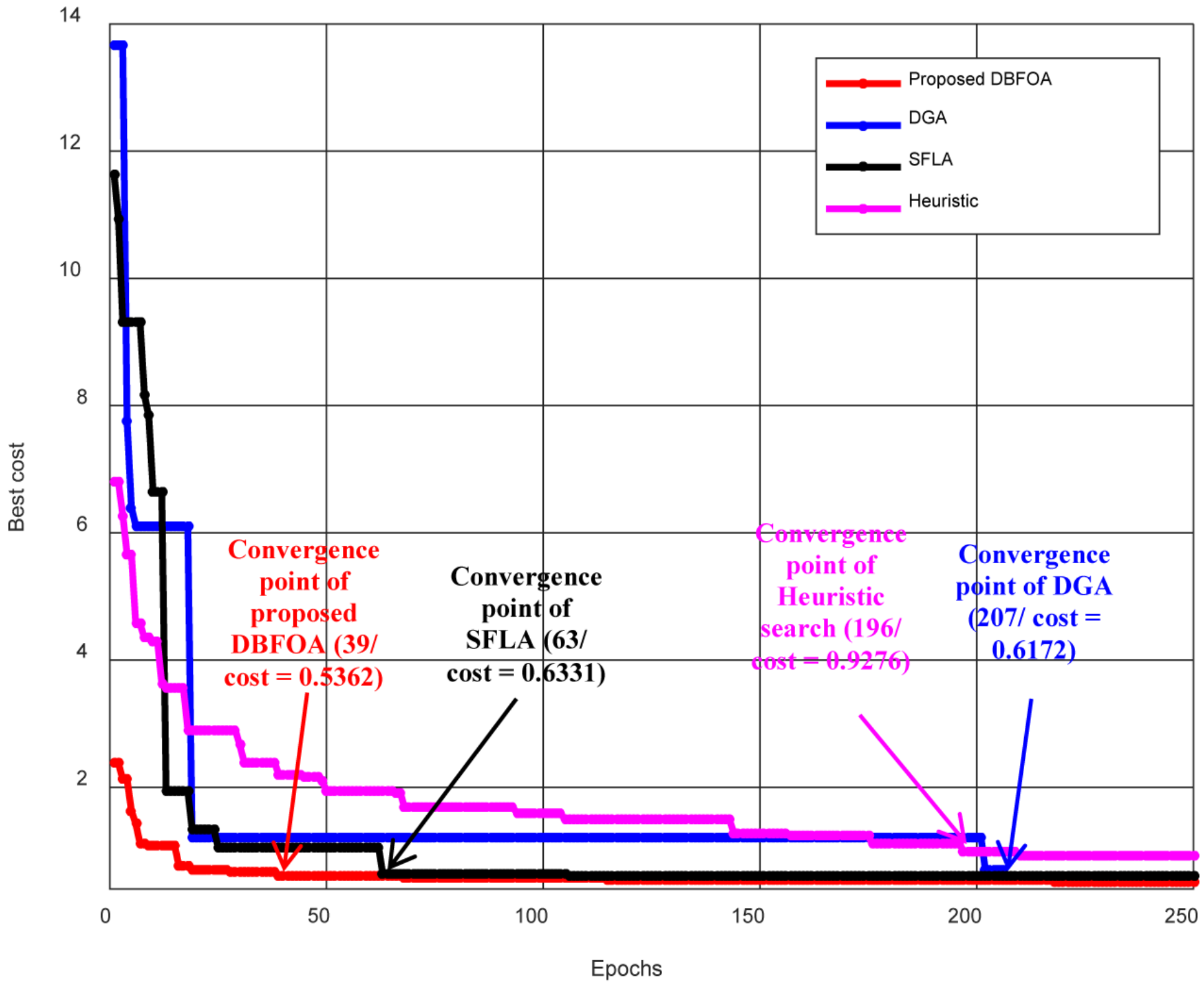

Figure 15: Performance comparison of proposed DBFOA with DGA, SFLA, and Heuristic search.

Table 7 shows the computational performance of the four algorithms. The average execution time of a single epoch for the three algorithms is almost the same. However, the proposed DBFOA converges to the optimal solution very fast compared to DGA, SFLA, and Heuristic Search.

Table 7: Computational efficiency of DBFOA, DGA, SFLA, and Heuristic search in terms of CPU time

| Algorithm | Execution time/Epoch (s) | Time to convergence/ (s) | Cost after 250 epochs |
|---|---|---|---|
| Proposed DBFOA | 0.888 | **33.744** | **0.5362** |
| DGA | 0.846 | 175.122 | 0.6172 |
| SFLA | 0.897 | 56.511 | 0.6331 |
| Heuristic search | 0.622 | 121.912 | 0.9276 |

# 7. Conclusion

In this paper, a novel method was introduced to mitigate the voltage unbalance in LV distribution grids through coordinated re-phasing of grid-connected rooftop PV systems. The optimum phase combination of grid-connected rooftop PV systems is determined from the modified discrete bacterial foraging optimization algorithm (DBFOA) at fixed time intervals. The DBFOA takes smart meter measurements such as load demands and PV generations as the inputs to determine the optimum phase configuration such that the resulting phase combination minimizes the overall voltage unbalance in the network, subject to various operating and network parameters. In order to perform automatic PV re-phasing, a PV re-phasing switch is introduced. This can connect to the output of the single-phase PV inverters to enable the PV re-phasing. To nullify the transient time associated with the automatic re-phasing switch, a half-bridge inverter arrangement is also proposed.

In order to demonstrate the effectiveness of the proposed re-phasing strategy, the PV re-phasing algorithm was simulated on a real LV distribution network. The time-varying nature of loads and solar PV was considered using the hourly load and PV generation profiles. The results show that the proposed re-phasing strategy can significantly reduce the voltage unbalance well below the 1% threshold line as compared to the fixed phase configuration during the daytime where PV penetration is high. Thereby the proposed PV re-phasing strategy facilitates utility providers to allow more rooftop solar systems into LV networks. The proposed case studies demonstrate that PV re-phasing technique can increase the renewable energy penetration into the considered LV network by 77%.

The main advantage of the proposed PV re-phasing technique compared to existing load and feeder reconfiguration techniques is that this method only deals with single-phase PV systems and hence, it will not create any impact on supply reliability. Furthermore, the initial cost for installation of PV re-phasing switches, and other indirect costs such as customer interruption cost and reliability cost are significantly minimal, as compare to existing state-of-the-art DFR and phase balancing techniques.

Even though the proposed re-phasing algorithm concentrates on mitigation of voltage unbalance, it is also possible to modify the objective function to minimize the number of PV re-phasings required for each PV re-phasing operation as well. Furthermore, additional research is required to determine the optimal timing for PV re-phasing operations and the effectiveness of a hybrid algorithm based on PV and load re-phasing techniques.

# 8. Acknowledgment

We would like to acknowledge the financial support provided by the National Science Foundation (NSF), Sri Lanka (research grant no: RG/2018/EA & ICT/01).

# 9. References

[1] Jadeja K. Major Technical issues with increased PV penetration on the existing electrical grid. Murdoch University, 2012. https://doi.org/https://core.ac.uk/download/pdf/11241964.pdf.

[2] Lu H, Zhao W. Effects of particle sizes and tilt angles on dust deposition characteristics of a ground-mounted solar photovoltaic system. Appl Energy 2018;220:514–26. https://doi.org/10.1016/j.apenergy.2018.03.095.

[3] Peters L, Madlener R. Economic evaluation of maintenance strategies for ground-mounted solar photovoltaic plants. Appl Energy 2017;199:264–80. https://doi.org/10.1016/j.apenergy.2017.04.060.

[4] Zhou Y, Chang F-J, Chang L-C, Lee W-D, Huang A, Xu C-Y, et al. An advanced complementary scheme of floating photovoltaic and hydropower generation flourishing

water-food-energy nexus synergies. Appl Energy 2020;275:115389. https://doi.org/10.1016/j.apenergy.2020.115389.

[5] Wang Q, Zhu Z, Wu G, Zhang X, Zheng H. Energy analysis and experimental verification of a solar freshwater self-produced ecological film floating on the sea. Appl Energy 2018;224:510–26. https://doi.org/10.1016/j.apenergy.2018.05.010.

[6] Hong T, Lee M, Koo C, Jeong K, Kim J. Development of a method for estimating the rooftop solar photovoltaic (PV) potential by analyzing the available rooftop area using Hillshade analysis. Appl Energy 2017;194:320–32. https://doi.org/10.1016/j.apenergy.2016.07.001.

[7] Cole W, Lewis H, Sigrin B, Margolis R. Interactions of rooftop PV deployment with the capacity expansion of the bulk power system. Appl Energy 2016;168:473–81. https://doi.org/10.1016/j.apenergy.2016.02.004.

[8] Carpinelli G, Mottola F, Proto D, Varilone P. Minimizing unbalances in low-voltage microgrids: Optimal scheduling of distributed resources. Appl Energy 2017;191:170–82. https://doi.org/10.1016/j.apenergy.2017.01.057.

[9] Zehir MA, Batman A, Sonmez MA, Font A, Tsiamitros D, Stimoniaris D, et al. Impacts of microgrids with renewables on secondary distribution networks. Appl Energy 2017;201:308–19. https://doi.org/10.1016/j.apenergy.2016.12.138.

[10] Wang L, Yan R, Saha TK. Voltage regulation challenges with unbalanced PV integration in low voltage distribution systems and the corresponding solution. Appl Energy 2019;256:113927. https://doi.org/10.1016/j.apenergy.2019.113927.

[11] Ma K, Li R, Hernando-Gil I, Li F. Quantification of Additional Reinforcement Cost from Severe Three-Phase Imbalance. IEEE Trans Power Syst 2017;32:4143–4. https://doi.org/10.1109/TPWRS.2016.2635383.

[12] Yaghoobi J, Islam M, Mithulananthan N. Analytical approach to assess the loadability of unbalanced distribution grid with rooftop PV units. Appl Energy 2018;211:358–67. https://doi.org/10.1016/j.apenergy.2017.11.030.

[13] Islam MR, Lu H, Hossain MJ, Li L. Mitigating unbalance using distributed network reconfiguration techniques in distributed power generation grids with services for electric vehicles: A review. J Clean Prod 2019;239:117932. https://doi.org/10.1016/j.jclepro.2019.117932.

[14] Fukami T, Onchi T, Naoe N, Hanaoka R. Compensation for neutral current harmonics in a three-phase four-wire system by a synchronous machine. IEMDC 2001. IEEE Int. Electr. Mach. Drives Conf. (Cat. No.01EX485), IEEE; n.d., p. 466–70. https://doi.org/10.1109/IEMDC.2001.939346.

[15] Singh B, Jayaprakash P, Kothari DP. A T-Connected Transformer and Three-leg VSC Based DSTATCOM for Power Quality Improvement. IEEE Trans Power Electron 2008;23:2710–8. https://doi.org/10.1109/TPEL.2008.2004273.

[16] Jayaprakash P, Singh B, Kothari DP. Three-Phase 4-Wire DSTATCOM Based on H-Bridge VSC with a Star/Hexagon Transformer for Power Quality Improvement. 2008 IEEE Reg.

10 Third Int. Conf. Ind. Inf. Syst., IEEE; 2008, p. 1–6. https://doi.org/10.1109/ICIINFS.2008.4798378.

[17] Jou H-L, Wu J-C, Wu K-D, Chiang W-J, Chen Y-H. Analysis of Zig-Zag Transformer Applying in the Three-Phase Four-Wire Distribution Power System. IEEE Trans Power Deliv 2005;20:1168–73. https://doi.org/10.1109/TPWRD.2005.844281.

[18] Enjeti P, Shireen W, Packebush P, Pitel I. Analysis and design of a new active power filter to cancel neutral current harmonics in three phase four wire electric distribution systems. Conf. Rec. 1993 IEEE Ind. Appl. Conf. Twenty-Eighth IAS Annu. Meet., IEEE; n.d., p. 939–46. https://doi.org/10.1109/IAS.1993.299011.

[19] Quinn CA, Mohan N. Active filtering of harmonic currents in three-phase, four-wire systems with three-phase and single-phase nonlinear loads. [Proceedings] APEC '92 Seventh Annu. Appl. Power Electron. Conf. Expo., IEEE; n.d., p. 829–36. https://doi.org/10.1109/APEC.1992.228328.

[20] Jou H-L, Wu K-D, Wu J-C, Li C-H, Huang M-S. Novel power converter topology for three-phase four-wire hybrid power filter. IET Power Electron 2008;1:164. https://doi.org/10.1049/iet-pel:20070171.

[21] Sreenivasarao D, Agarwal P, Das B. Neutral current compensation in three-phase, four-wire systems: A review. Electr Power Syst Res 2012;86:170–80. https://doi.org/10.1016/j.epsr.2011.12.014.

[22] Ramos-figueroa O, Quiroz-castellanos M, Mezura-montes E. Metaheuristics to solve grouping problems : A review and a case study. Swarm Evol Comput 2020;53:100643. https://doi.org/10.1016/j.swevo.2019.100643.

[23] Kashem MA, Ganapathy V, Jasmon GB. Network reconfiguration for load balancing in distribution networks. IEE Proc - Gener Transm Distrib 1999;146:563. https://doi.org/10.1049/ip-gtd:19990694.

[24] Babu PR, Kumar KA, Teja GC. New heuristic search approach to enhance the distribution system load balance. 2013 Int. Conf. Power, Energy Control, IEEE; 2013, p. 159–64. https://doi.org/10.1109/ICPEC.2013.6527642.

[25] Babu PR, Shenoy R, Ramya N, Soujanya, Shetty S. Implementation of ACO technique for load balancing through reconfiguration in electrical distribution system. 2014 Annu. Int. Conf. Emerg. Res. Areas Magn. Mach. Drives, IEEE; 2014, p. 1–5. https://doi.org/10.1109/AICERA.2014.6908233.

[26] Yuehao Y, Zhongqing Z, Wei B, Jun X, Limin Q, Yaoheng D. Optimal distribution network reconfiguration for load balancing. 2016 China Int. Conf. Electr. Distrib., vol. 2016- Septe, IEEE; 2016, p. 1–4. https://doi.org/10.1109/CICED.2016.7576313.

[27] Fu-Yuan Hsu, Men-Shen Tsai. A Multi-Objective Evolution Programming Method for Feeder Reconfiguration of Power Distribution System. Proc. 13th Int. Conf. on, Intell. Syst. Appl. to Power Syst., vol. 2005, IEEE; 2005, p. 55–60. https://doi.org/10.1109/ISAP.2005.1599241.

[28] Taher SA, Karimi MH. Optimal reconfiguration and DG allocation in balanced and

unbalanced distribution systems. Ain Shams Eng J 2014;5:735–49. https://doi.org/10.1016/j.asej.2014.03.009.

[29] Vitor TS, Vieira JCM. Optimal voltage regulation in distribution systems with unbalanced loads and distributed generation. 2016 IEEE Innov. Smart Grid Technol. - Asia, IEEE; 2016, p. 942–7. https://doi.org/10.1109/ISGT-Asia.2016.7796512.

[30] Nara K, Mishima Y, Satoh T. Network reconfiguration for loss minimization and load balancing. 2003 IEEE Power Eng. Soc. Gen. Meet. (IEEE Cat. No.03CH37491), vol. 4, IEEE; 2003, p. 2413–8. https://doi.org/10.1109/PES.2003.1271019.

[31] Qin Zhou, Shirmohammadi D, Liu W-HE. Distribution feeder reconfiguration for service restoration and load balancing. IEEE Trans Power Syst 1997;12:724–9. https://doi.org/10.1109/59.589664.

[32] Tolabi HB, Ali MH, Shahrin Bin Md Ayob, Rizwan M. Novel hybrid fuzzy-Bees algorithm for optimal feeder multi-objective reconfiguration by considering multiple-distributed generation. Energy 2014;71:507–15. https://doi.org/10.1016/j.energy.2014.04.099.

[33] Ke Y-L, Chen C-S, Kang M-S, Wu J-S, Lee T-E. Power Distribution System Switching Operation Scheduling for Load Balancing by Using Colored Petri Nets. IEEE Trans Power Syst 2004;19:629–35. https://doi.org/10.1109/TPWRS.2003.821433.

[34] Lin C-H. Distribution network reconfiguration for load balancing with a coloured Petri net algorithm. IEE Proc - Gener Transm Distrib 2003;150:317. https://doi.org/10.1049/ip-gtd:20030199.

[35] Ji H, Wang C, Li P, Zhao J, Song G, Ding F, et al. An enhanced SOCP-based method for feeder load balancing using the multi-terminal soft open point in active distribution networks. Appl Energy 2017;208:986–95. https://doi.org/10.1016/j.apenergy.2017.09.051.

[36] Zhai HF, Yang M, Chen B, Kang N. Dynamic reconfiguration of three-phase unbalanced distribution networks. Int J Electr Power Energy Syst 2018;99:1–10. https://doi.org/10.1016/j.ijepes.2017.12.027.

[37] Santos SF, Fitiwi DZ, Cruz MRM, Cabrita CMP, Catalão JPS. Impacts of optimal energy storage deployment and network reconfiguration on renewable integration level in distribution systems. Appl Energy 2017;185:44–55. https://doi.org/10.1016/j.apenergy.2016.10.053.

[38] Kaveh MR, Hooshmand RA, Madani SM. Simultaneous optimization of re-phasing, reconfiguration and DG placement in distribution networks using BF-SD algorithm. Appl Soft Comput J 2018;62:1044–55. https://doi.org/10.1016/j.asoc.2017.09.041.

[39] Soltani SH, Rashidinejad M, Abdollahi A. Dynamic phase balancing in the smart distribution networks. Int J Electr Power Energy Syst 2017;93:374–83. https://doi.org/10.1016/j.ijepes.2017.06.016.

[40] Kuo C-C, Chao Y-T. Energy management based on AM/FM/GIS for phase balancing application on distribution systems. Energy Convers Manag 2010;51:485–92. https://doi.org/10.1016/j.enconman.2009.10.011.

[41] Jinxiang Zhu, Mo-Yuen Chow, Fan Zhang. Phase balancing using mixed-integer programming [distribution feeders]. IEEE Trans Power Syst 1998;13:1487–92. https://doi.org/10.1109/59.736295.

[42] Chitra R, Neelaveni R. A realistic approach for reduction of energy losses in low voltage distribution network. Int J Electr Power Energy Syst 2011;33:377–84. https://doi.org/10.1016/j.ijepes.2010.08.033.

[43] Hooshmand RA, Soltani S. Fuzzy Optimal Phase Balancing of Radial and Meshed Distribution Networks Using BF-PSO Algorithm. IEEE Trans Power Syst 2012;27:47–57. https://doi.org/10.1109/TPWRS.2011.2167991.

[44] Das CK, Bass O, Kothapalli G, Mahmoud TS, Habibi D. Optimal placement of distributed energy storage systems in distribution networks using artificial bee colony algorithm. Appl Energy 2018;232:212–28. https://doi.org/10.1016/j.apenergy.2018.07.100.

[45] Zhu J, Bilbro G, Mo-Yuen Chow. Phase balancing using simulated annealing. IEEE Trans Power Syst 1999;14:1508–13. https://doi.org/10.1109/59.801943.

[46] Shahnia F, Ghosh A, Ledwich G, Zare F. Voltage unbalance improvement in low voltage residential feeders with rooftop PVs using custom power devices. Int J Electr Power Energy Syst 2014;55:362–77. https://doi.org/10.1016/j.ijepes.2013.09.018.

[47] Gray MK, Morsi WG. Economic assessment of phase reconfiguration to mitigate the unbalance due to plug-in electric vehicles charging. Electr Power Syst Res 2016;140:329–36. https://doi.org/10.1016/j.epsr.2016.06.008.

[48] Dinesh C, Welikala S, Liyanage Y, Ekanayake MPB, Godaliyadda RI, Ekanayake J. Non-intrusive load monitoring under residential solar power influx. Appl Energy 2017;205:1068–80. https://doi.org/10.1016/j.apenergy.2017.08.094.

[49] Welikala S, Thelasingha N, Akram M, Ekanayake PB, Godaliyadda RI, Ekanayake JB. Implementation of a robust real-time non-intrusive load monitoring solution. Appl Energy 2019;238:1519–29. https://doi.org/10.1016/j.apenergy.2019.01.167.

[50] Wen H, Cheng D, Teng Z, Guo S, Li F. Approximate Algorithm for Fast Calculating Voltage Unbalance Factor of Three-Phase Power System. IEEE Trans Ind Informatics 2014;10:1799–805. https://doi.org/10.1109/TII.2014.2327485.

[51] K.M. Passino. Biomimicry of bacterial foraging for distributed optimization and control. IEEE Control Syst 2002;22:52–67. https://doi.org/10.1109/MCS.2002.1004010.

[52] Das S, Biswas A, Dasgupta S, Abraham A. Bacterial Foraging Optimization Algorithm: Theoretical Foundations, Analysis, and Applications. Found. Comput. Intell. Vol. 3. Stud. Comput. Intell., Berlin, Heidelberg: Springer-Verlag; 2009, p. 23–55. https://doi.org/10.1007/978-3-642-01085-9_2.

[53] Devi S, Geethanjali M. Application of Modified Bacterial Foraging Optimization algorithm for optimal placement and sizing of Distributed Generation. Expert Syst Appl 2014;41:2772–81. https://doi.org/10.1016/j.eswa.2013.10.010.

[54] Devabalaji KR, Ravi K, Kothari DP. Optimal location and sizing of capacitor placement in

radial distribution system using Bacterial Foraging Optimization Algorithm. Int J Electr Power Energy Syst 2015;71:383–90. https://doi.org/10.1016/j.ijepes.2015.03.008.

[55] Sathish Kumar K, Jayabarathi T. Power system reconfiguration and loss minimization for an distribution systems using bacterial foraging optimization algorithm. Int J Electr Power Energy Syst 2012;36:13–7. https://doi.org/10.1016/j.ijepes.2011.10.016.

[56] Hooshmand R-A, Parastegari M, Morshed MJ. Emission, reserve and economic load dispatch problem with non-smooth and non-convex cost functions using the hybrid bacterial foraging-Nelder–Mead algorithm. Appl Energy 2012;89:443–53. https://doi.org/10.1016/j.apenergy.2011.08.010.